\documentclass[a4paper,11pt]{article}
\pdfoutput=1
\usepackage{jcappub} 
\usepackage[T1]{fontenc}
\usepackage{amsfonts}                  
\usepackage{slashed}                   
\usepackage{bm}                        
\usepackage{array}                      
\usepackage{multirow}                   
\usepackage{graphicx}                   
\usepackage{hhline}                     
\usepackage{threeparttable}
\usepackage{pgfplots}
\usepackage{ esint }
 \usepackage{ amssymb }
\usepackage{hyperref}
 \usepackage{ textcomp }
 \usepackage{graphicx}
 \usepackage{ dsfont }
 \usepackage{mathtools}
 \usepackage{tocloft}
 \usepackage{slashed}
 \usepackage{amsmath}
 \usepackage{mathrsfs}
 \usepackage{lettrine}
 \usepackage{relsize}

\title{ Asymmetric Dark Matter and CP Violating Scatterings in a UV Complete Model}

\author[]{Iason Baldes,}
\author[]{Nicole F. Bell,}
\author[]{Alexander J. Millar}
\author[]{and Raymond R. Volkas}

\affiliation[]{ARC Centre of Excellence for Particle Physics at the Terascale,\\
School of Physics, The University of Melbourne, Victoria 3010, Australia}

\emailAdd{i.baldes@student.unimelb.edu.au}
\emailAdd{n.bell@unimelb.edu.au}
\emailAdd{amillar@student.unimelb.edu.au}
\emailAdd{raymondv@unimelb.edu.au}

\abstract{We explore possible asymmetric dark matter models using CP violating scatterings to generate an asymmetry. In particular, we introduce a new model, based on DM fields coupling to the SM Higgs and lepton doublets, a neutrino portal, and explore its UV completions. We study the CP violation and asymmetry formation of this model, to demonstrate that it is capable of producing the correct abundance of dark matter and the observed matter-antimatter asymmetry. Crucial to achieving this is the introduction of interactions which violate CP with a $T^{2}$ dependence.}

\begin{document}
\maketitle
\flushbottom

\section{Introduction}
Measurements of the CMB and BBN show that the baryon-to-entropy density ratio in our universe~\cite{Hinshaw:2012aka,Ade:2013zuv},
	\begin{equation}
	Y_B=(0.86\pm 0.01)\times 10^{-10},
	\end{equation}
 is much higher than the SM predicts~\cite{Zeldovich,PhysRevLett.17.712}. Further, observations show us that the Standard Model (SM) only describes 15\% of the matter in the universe, the remainder being dark matter (DM). In light of this, it is very suggestive that the mass densities of dark and visible matter are so similar, with $\Omega_{DM} \simeq5\Omega_{VM}$ where $\Omega_{DM}$ ($\Omega_{VM}$) is the DM (visible matter) density expressed as a ratio to the critical density. In many standard DM scenarios this must be taken as a coincidence. On the other hand, in asymmetric dark matter (ADM) models the DM density is due to an asymmetry which is linked to the baryon asymmetry in some way~\cite{Davoudiasl:2012uw,Petraki:2013wwa,Zurek:2013wia}.\\

To do this requires that the DM and visible sectors be related - possibly at very high energies. Suppose there is a conserved global symmetry between dark and visible matter such that $B-L-D$ is conserved, where $B$ and $L$ are the usual baryon and lepton numbers and $D$ is the dark matter charge. For asymmetries to form, $B-L+D$  must be broken. Then if $\Omega_{DM}$ is due to a $D$ asymmetry, the DM mass must lie in a narrow range to ensure the correct abundance~\cite{Petraki:2013wwa},
\begin{equation}
m_{DM} \simeq Q_D\times  (1.7 - 5)   \; \textrm{GeV} \label{dm},
\end{equation}
where $Q_D$ is the charge of dark matter under $D$. The value of 1.7 GeV holds for totally asymmetric dark matter with $Q_D=1$ where baryogenesis occurs entirely before the electroweak phase transition (EWPT), as will be considered in this work.\\ 

As in normal baryogenesis, a process must satisfy the Sakharov conditions (particle number violation, C and CP violation and thermal non-equilibrium), to be capable, in principle, of forming a particle number asymmetry~\cite{Sakharov:1967dj}. Decays of heavy particles, first order phase transitions and the Affleck-Dine mechanism have all been proposed as mechanisms to create a baryon asymmetry and related DM asymmetry. A full list of references can be found in reviews~\cite{Davoudiasl:2012uw,Petraki:2013wwa,Zurek:2013wia}. In this paper we will instead study the creation of an asymmetry through CP violating scatterings. \\

The possibility of using CP violating scatterings or coannihilations to generate ADM was first proposed in~\cite{Bento:2001rc}, the idea of using asymmetric freeze-out was introduced in~\cite{Farrar:2005zd} and baryogenesis via annihilating particles was also studied in~\cite{Gu:2009yx}. The details of such a mechanism in ADM were further explored in a toy model context in~\cite{Baldes:2014gca}. A baryogenesis mechanism, using the neutron portal effective operator, in which CP violating scatterings typically dominate over decays, was discussed by the present authors in~\cite{Baldes:2014rda}. CP violating scatterings have also been studied in leptogenesis, in which they are negligible compared to decays outside of the weak washout regime~\cite{Nardi:2007jp,Fong:2010bh,Fong:2013wr}. The aim of the current paper is to explore for the first time viable and UV complete ADM models which use the mechanism of CP violating scatterings during freeze-out. The DM will be stabilised using a $\mathds{Z}_{2}$ symmetry rather than couplings with a strong temperature dependence as in~\cite{Farrar:2005zd}.\\

CP violating $2 \leftrightarrow 2$ processes are also crucial to the WIMPy baryogenesis scenarios, in which the baryon asymmetry is generated by the annihilations of weakly interacting massive particles (WIMPs) during freeze-out. However, the DM density is not due to an asymmetry in those scenarios~\cite{Cui:2011ab,Bernal:2012gv,Bernal:2013bga,Kumar:2013uca,Racker:2014uga}. CP violating scattering has also been studied in an ADM context, albeit for freeze-in scenarios~\cite{Bento:2001rc,Hook:2011tk,Unwin:2014poa}. \\

The structure of this paper is as follows. In Section \ref{alternatives} we introduce the neutrino portal. We couple $\overline{L}H$ to an exotic scalar and fermion, the lightest of which will form DM, and discuss possible UV completions. In Section \ref{findingasym} we analyse asymmetry production in the minimal inert two-Higgs-doublet UV completion of the neutrino portal. We find it cannot produce an asymmetry of the observed size. Due to this we discuss a simple extension in Section \ref{leptonportalext}, introducing an additional Higgs doublet, and show the observed asymmetry can be obtained due to the additional CP violation present. We then conclude. Alternative operators are discussed in Appendices \ref{neutronportal} and \ref{otherops}.

\label{sec:intro}

\section{The Neutrino Portal}
\label{alternatives}
When trying to write down a model of ADM, it is simplest to start with the lowest mass dimension gauge invariant operator involving lepton or baryon number, $\overline L H$ ($H\sim(1,2,-1)$). This must be paired with at least two non-SM particles in order to stabilise dark matter with a symmetry, leading to a neutrino portal~\cite{Falkowski:2009yz,Macias:2015cna}. This can be achieved via the effective operator
\begin{equation} 
	g \overline L Y H \phi\label{efflep},
\end{equation}
where $g$ is a coupling with inverse dimension of mass, $L$ ($H$) is the SM Lepton (Higgs) doublet and $Y$ ($\phi$) is an exotic fermion (scalar). The latter two fields will comprise our dark sector, in which the lightest particle's stability will be ensured by a $\mathds{Z}_2$ symmetry. Both $\phi$ and $Y$ are odd under this symmetry, and all SM particles are even.\\

It is preferable to have only one dark sector particle carry a baryon number equivalent $D$, because if both particles did, then they must both be stable. We can define a dark baryon number $D=(N_{\phi}+N_{Y})/2$ (where $N_X$ is the particle number of $X$), so that $B-L-D$ is conserved and $B-L+D$ is broken, as desired for ADM. Unfortunately if both $\phi~\text{and}~Y$ carry $D$ we have a U(1) symmetry in addition to $B-L-D$ conservation. This is obtained by rephasing only the dark sector particles, giving
\begin{equation}
\Delta \phi-\Delta Y=0, 
\label{twodirac}
\end{equation}
where $\Delta X =N_X-N_{\overline X}$. If $\phi$ or $Y$ decays, \eqref{twodirac} requires that the asymmetry stored in the remaining particle vanishes.\footnote{If there are additional dark sector particles carrying $D$, then an extension of this argument still holds: at least two must be stable. As this is non-minimal, we do not consider that scenario in this work} Then baryogenesis does not occur as $B-L=0$.\footnote{As a special case this can be avoided if the scatterings freeze-out before the EWPT but the particle decays after the EWPT, because $B$ and $L$ are no longer related by electroweak sphalerons.} This problem is avoided if only one dark sector particle carries $D$, e.g. $D=N_{\phi}$.  \\

To keep both $\phi$ and $Y$ stable would require kinematically disallowing decays with a rather implausible mass difference smaller than the neutrino masses. This mass difference cannot be obtained as one particle species must go out-of-equilibruim above the EWPT (so sphalerons can reprocess $\Delta L$ into $\Delta B$) and DM must be of order a few GeV. While it is possible that additional interactions could prevent $Y$ or $\phi$ from carrying an extra dark conserved charge, for simplicity we will not consider that scenario. This leaves two possibilities: either $\phi$ is a complex scalar and $Y$ a Majorana fermion, or $\phi$ is a real scalar and $Y$ a Dirac fermion. We choose the former case as it leads to more CP violating phases and permits a simpler method to annihilate the symmetric component. Asymmetry production should be the same in both cases, though they will have phenomenologically different dark matter (scalar versus fermion). \\

\begin{figure}[t!]
\centering
\includegraphics[width=350pt]{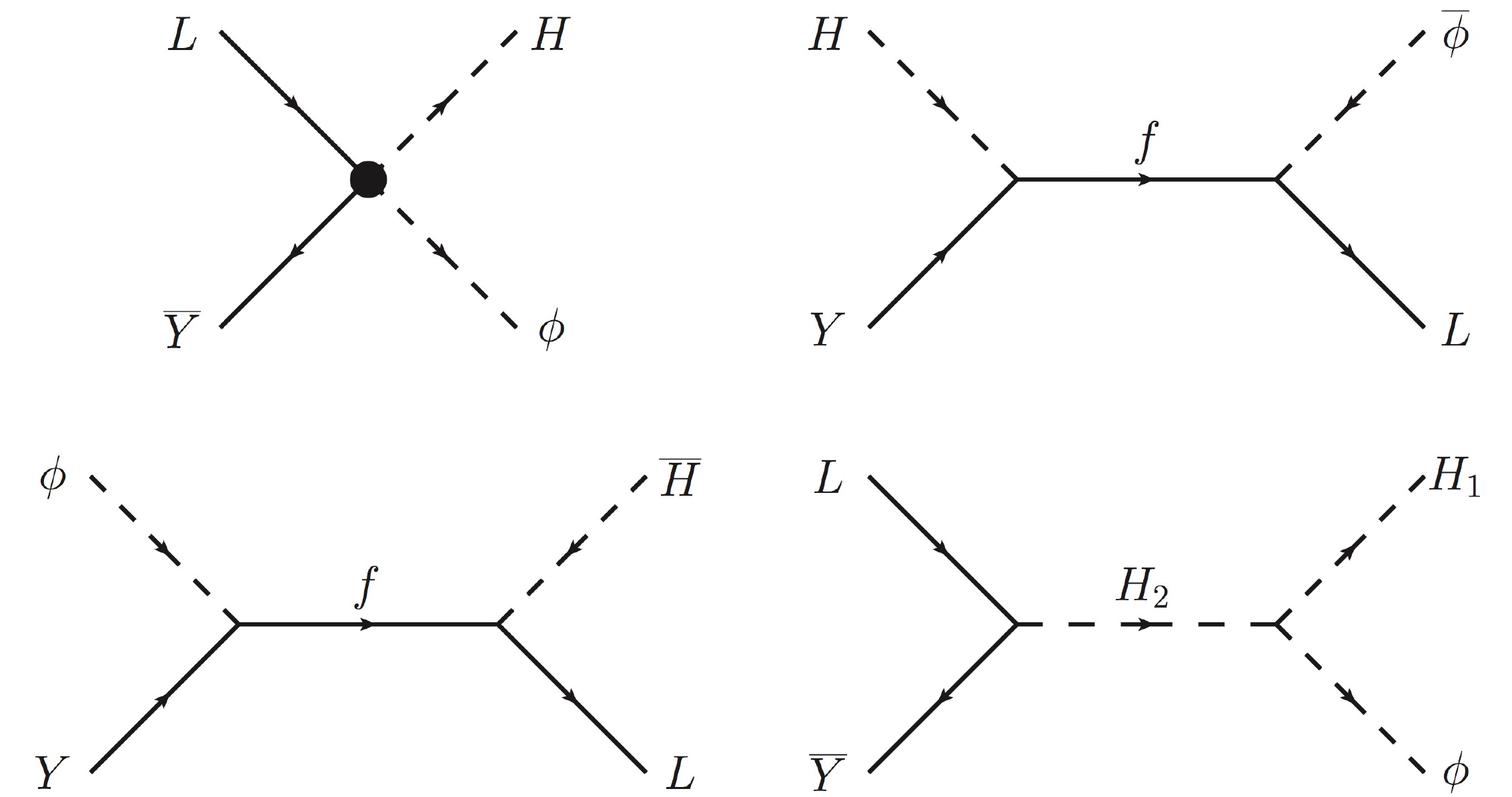} 
\caption{The interaction $\overline L Y H \phi$ together with its UV completions. We label them as case 1 (top right), case 2 (bottom left) and case 3 (bottom right).}
\label{fig:5}
\end{figure}

We envision the $Y$s as the most massive particles (barring the mediator), going out-of-equilibrium before our dark matter candidate $\phi$. As they do not store an asymmetry, they can safely decay without removing the asymmetry generated by the scatterings. To satisfy unitarity constraints, which we will discuss in Section \ref{notation}, there must be two generations of $Y$s and one of $\phi$. Our ADM will be $\phi$, which carries a non-zero baryon number equivalent $D=N_{\phi}$, preserving $B-L-D$. \\

In this paper we consider UV completions of the effective interaction, eq.~(\ref{efflep}). The simplest tree level possibilities are depicted diagrammatically in fig.~\ref{fig:5}. The thermal histories of all UV completions are essentially the same. At high temperature the $Y$s are kept in equilibrium by the rapid $2\to 2$ interactions. These interactions eventually freeze-out and the $Y$s depart from thermal equilibrium, satisfying the third Sakharov condition. As these scatterings violate CP, a non-zero $B-L$ is generated along with an associated $D$ asymmetry stored in the $\phi$, creating a baryon asymmetry via the sphalerons (as long as this occurs above the EWPT). Subsequently, the $Y$s decay into leptons and $\phi$, potentially leading to further asymmetry formation. To see which process dominates, as well as the size of the asymmetry formed, it is necessary to solve the Boltzmann equations, which we will do in Section \ref{findingasym}. The three simple UV completions are as follows.
\begin{itemize}

\item The first case requires the mediator, $f$, to be a vector-like lepton. The $\mathds{Z}_2$ symmetry prevents $f$ from Yukawa coupling via the Higgs to the SM leptons, avoiding most of the limits on vector-like leptons~\cite{Ishiwata:2013gma,Falkowski:2013jya}. For CP violation to arise at one-loop it is necessary to include two copies of $f$.

\item The second case has a Dirac SM singlet, $f$ as intermediary. The $\mathds{Z}_2$ symmetry precludes the $Y$ from acting as an intermediate particle and coupling to the leptons, which would destroy the global $B-L-D$. If $Y$ is Dirac, rather than Majorana, this operator is very similar to~\cite{Bento:2001rc}. In~\cite{Bento:2001rc}, a Majorana fermion mediates an interaction between $\overline L H$ and a mirror copy, $\overline{L^{'}}H^{'}$, using different initial temperatures to create out-of-equilibrium conditions. CP violation is the same as in the first completion.

\item The third case is an extension of inert two-Higgs-doublet models (IDM) \cite{Barbieri:2006dq}. The intermediate particle, $H_2$, has the same quantum numbers as the SM Higgs but cannot play the same role. While CP violation is possible with just one inert Higgs doublet, we will show that to get a sufficient asymmetry to form, two inert Higgs doublets are necessary. While regular IDMs do have a dark matter candidate, they do not explain the ratio of dark and visible matter; the mass ranges required for $H_2$ to be ADM instead of $\phi$ are excluded by a combination of collider searches, electroweak precision tests and direct detection constraints \cite{LopezHonorez:2006gr,Dhen:2015wra}.\footnote{An exception to this can be found in~\cite{Dhen:2015wra}, where an inert Higgs transfers a pre-existing asymmetry between the dark and visible sectors.}

\end{itemize}
As the IDM completion is particularly illustrative we will study this case in more detail. 

\subsection{Inert Higgs Doublet Completion}
\label{leptonport}
To see a specific model in action, we will study the IDM completion of the neutrino portal.
As in IDM, we add a massive scalar $SU(2)_L$ doublet which will function as our heavy intermediary. We also introduce a new parity, with the Lagrangian symmetric under $H_2\rightarrow -H_2$, $Y \rightarrow -Y$, and $\phi \rightarrow -\phi$. All SM particles are even under this parity. We call $H_2$ inert as it does not acquire a non-zero vacuum expectation value (VEV), and does not play a role in the fermion mass generation. Where this work differs from traditional IDMs is that in our case $H_2$ carries a non-zero $B-L$.\footnote{The main effect of this is on neutrino masses. In standard IDMs neutrinos gain Majorana masses through radiative corrections \cite{Ma:2006km}, which requires a term $\lambda_5(H_2^{*}H_1)^2$. As this term breaks $B-L-D$, for simplicity we consider instead neutrinos gaining a mass through the Higgs mechanism (with the addition of right handed neutrinos even under the $\mathds{Z}_2$). This can be a Dirac mass or a Majorana mass if lepton number is softly broken.} The relevant additions to the SM Lagrangian are
\begin{equation}
\Delta \mathcal L = - \lambda_{ia}H_2\overline L_i Y_a-\kappa H_1 H_2^{\dagger} \phi - \lambda_1 |H_1|^2 |H_2|^2 -\lambda_2 |H_1^{\dagger}H_2|^2 -\lambda_3 |H_1|^2|\phi|^2-\lambda_4 |\phi|^4+H.c. \label{minlang}
\end{equation}
This induces \eqref{efflep} after integrating out the mediating $H_2$. After considering field rephasings, there are three physical CP violating phases, that we write as
\begin{align}
\theta_1= \ &\operatorname{Arg}(\lambda_{12}),\\
\theta_2= \ &\operatorname{Arg}(\lambda_{22}),\\
\theta_3= \ &\operatorname{Arg}(\lambda_{32}),
\end{align}
with all other couplings real (without loss of generality).
For our example solutions we choose $\theta_1=\theta_2=\theta_3=\pi/4$.\\

We consider the $Y$ to have a mass higher than the scale of the EWPT, and $H_2$ to be heavy enough to be considered always off shell, typically several orders of magnitude heavier than the next lightest particle, $Y_2$.  As the neutrino portal does not violate baryon number directly, electroweak sphalerons are required to reprocess the lepton asymmetry into a baryon asymmetry. For the parameters considered in this paper, all relevant processes finish before the EWPT. \\

We assume that there is an additional process that keeps the $\phi$ in equilibrium and annihilates the symmetric component, but remain agnostic as to the details of this process. For an example, $\phi$ quartic coupling to a light real scalar field could be used for this purpose. In more baroque extensions it is also possible for $\phi$ to decay into lighter particles, and for them to have new interactions that annihilate their symmetric component. While for the purposes of this paper the exact method is irrelevant, different mechanisms will provide different detection prospects. For example, if $\phi$ annihilates into a light, stable state then there is a significant contribution to dark radiation~\cite{Petraki:2013wwa}.\\
 
Originally it was thought possible to use CP violating scatterings to simultaneously create an asymmetry and annihilate the symmetric component \cite{Farrar:2005zd}. Unfortunately, this is impossible as it would require the freeze-out temperature to be of order $m_{\text{DM}}/25$. For operators like those of \cite{Farrar:2005zd}, decays prevent an asymmetry forming at arbitrarily low freeze-out temperatures. If we make the coupling strength large enough to delay scattering freeze-out until $T_{\text{freeze-out}}\sim m_{\text{DM}}/25$, the lifetime becomes short enough that decays and inverse decays keep the particle in equilibruim, so Sakharov's conditions are never met. We find numerically that the lifetime becomes shorter than $t_{\text{freeze-out}}$ well before $T_{\text{freeze-out}}\sim m_{\text{DM}}/25$. As discussed in Appendix \ref{whataboutsym}, models that might have this feature are disfavoured by limits on non-SM coloured particles.\\

\subsection{Phenomenological Constraints}
Due to the mass range of $H_2$ we are considering most of the couplings related to asymmetry formation in our theory are unconstrained, but there are still some restrictions from electroweak considerations.\footnote{While directly searching the full possible mass range of $H_2$ is impossible at the LHC, it is possible to rule out high scale baryogenesis in general. If low scale lepton number violation (LNV) is observed, either through direct observation of LNV  at the LHC or through a combination of neutrinoless double beta decay and lepton flavour violation, then the washout induced by this LNV makes any high scale baryogenesis irrelevant \cite{Deppisch:2013jxa,Deppisch:2015yqa}.}

\subsubsection{Electroweak Phase Transition} 
While the population of $H_2$ is negligible during the EWPT, it is still worth considering the effects of the EWPT on $H_2$ as it couples to $\phi$. In addition, while $\phi$ is a SM singlet it can still potentially influence the EWPT. During electroweak symmetry breaking (EWSB), $H_1$ acquires a VEV $v$ and acts as the SM Higgs; on the other hand $H_2$ only experiences mass splitting. It is possible to parameterize $H_2$ as 
\begin{equation}
H_2=\binom{H_2^0}{H_2^{-}},
\end{equation}
where $H^{0}$ and $H^-$ are complex scalars. After EWSB we have 
\begin{align}
m_{H_2^{0}}^2 &= m_{H_2}^2+(\lambda_1+\lambda_2) v^2,\\
m_{H_2^{-}}^2 &= m_{H_2}^2+\lambda_1 v^2.
\end{align}
There is also a contribution to the mass of $\phi$ when $H_1$ acquires a VEV, as well as mass mixing between $H_2^0$ and $\phi$. This gives us a mass matrix in the $(\phi,H_2^0)$ basis
\begin{equation}
\left( \begin{array}{ccc}
 m_{\phi}^2+\lambda_3 v^2 & \kappa v \\
\kappa v & m_{H_2^0}^2
 \end{array} \right).
\end{equation}
We expect this to diagonalise to a heavy state and a light state. We will label the light state $\phi^{'}$ as the dark matter admixture is mostly $\phi$, with mixing angle $\kappa v/m_{H_2^0}^2$. We have a light enough dark matter candidate for small mixing angles,
 \begin{equation}
 v \kappa /m_{H_2^0} \lesssim (m_{\phi}^2+\lambda_3 v^2)^{1/2}. \label{mixing}
 \end{equation}
This requirement for small mixing is the only real constraint relevant to asymmetry formation. For $\kappa \simeq m_Y$ and $m_{\phi} \simeq 0$, we have $m_{\phi^{'}}^2 \simeq \lambda_3 v^2$.
As we envision $\phi^{'}$ to be ADM, we must have $\lambda_3 \lesssim 5 \times 10^{-5}$ to get the correct relic density ($m_{\phi^{'}}\simeq 1.7$ GeV as in \eqref{dm}).\footnote{In this case $\lambda_3$ is much too small for $\phi$ to be detected at the LHC via invisible Higgs decays. However, if there are significant cancelations between $m_{\phi}^2$ and $\lambda_3 v^2$ it is possible for $\lambda_3$ to be large enough for the invisible decays of the SM Higgs into $\phi$ to be detectable at the LHC, though this is not required for our scenario to work~\cite{Sage:2015wfa}.} \\

Introducing $\phi$ can potentially alter the EWPT, but only via its quartic coupling to $H_1$. One might worry that if the EWPT is made first order, there would be two competing methods for baryogenesis - electroweak baryogenesis and scatterings. From the analysis in \cite{Espinosa:1993bs} it can be shown that the EWPT is first order if 
\begin{equation}
\frac{\lambda_3^4}{128\pi^2}-\frac{1}{3}\lambda_{H_1}^2(\lambda_3+\lambda_4)> \left[\lambda_{H_1}\left(\frac{\lambda_3}{6}+\frac{y_t}{2}\right)-\frac{\lambda_3^3}{32\pi^2}\right]\left(\frac{m_{\phi}}{v}\right)^2,
\end{equation}
 where $\lambda_{H_1}$ is the quartic self coupling of $H_1$ and $y_t$ is the Yukawa coupling of $H_1$ to the top quark. As $\lambda_3$ is small, the EWPT remains second order. We make the approximation that the electroweak phase transition occurs the same way as in the SM.

\subsubsection{Electroweak Precision Tests}
As $H_2$ still couples to the electroweak gauge bosons, electroweak precision tests are, in principle, sensitive to $H_2$. More precisely the electroweak precision tests are only sensitive to the mass splitting of $H_2$ \cite{Barbieri:2006dq}. Standard electroweak precisions tests can be cast in terms of the "oblique" or Peskin-Takeuchi parameters S, T and U which parametrize the radiative corrections due to new physics~\cite{Peskin:1991sw}. From \cite{Barbieri:2006dq} the main contributions to electroweak precision tests (at one loop order) can be written as
\begin{align}
\Delta T 
&\simeq \frac{\lambda_2}{32\pi^2\alpha}\simeq 0.4\lambda_2,\\
\Delta S&\simeq\frac{\lambda_2 v^2}{8\pi^2m_{H_2^{-}}^2}\simeq 8\times 10^{-6} \times\lambda_2\left(\frac{10~\text{TeV}}{m_{H_2^{-}}^2}\right).
\end{align}
The experimental values are $\Delta T_{U=0}=0.08\pm0.07$ and $\Delta S_{U=0}=0.05\pm0.09$, so $\lambda_2 \lesssim 0.2$ is required to satisfy electroweak precision tests \cite{Baak:2012kk}. Fortunately, $\lambda_2$ does not affect asymmetry production, so this constraint does not seriously affect the model. $\Delta S$ is negligible for all sensible parameter choices. 
\subsubsection{Self Interactions}
While the self couplings of $\phi$ do not affect asymmetry production directly, they will contribute to the thermal mass of $\phi$.
The most stringent limits on dark matter self interactions come from the Bullet Cluster and similar colliding clusters. From this there is the constraint,  \cite{Harvey:2015hha,Randall:2007ph,Kouvaris:2011fi}
\begin{equation}
\lambda_{\text{4}}^2\lesssim 4\times 10^{5}\left(\frac{m_{\phi^{'}}}{1~\text{GeV}}\right)^3.
\end{equation}
This leaves $\lambda_4$ essentially unconstrained. As the thermal mass of $\phi$ will be determined by $\lambda_4$, this freedom is useful for the kinematics. We use the thermal masses described in Appendix~\ref{ThermalMasses} throughout this paper.\\

Fortuitously only one of these constraints affect parameters required for asymmetry formation, so we are free to choose parameters to maximise the asymmetry formed. The only real constraint is \eqref{mixing}.

\section{Finding the Asymmetry}
\label{findingasym}
\subsection{Notation and Tools}
 \label{notation}
Now that we have invented some interactions satisfying the Sakharov conditions, we need to know the magnitude and nature of the asymmetry they generate. Our main technique for calculating the evolution of an asymmetry will be Boltzmann equations. We adopt the same $W(\psi,a\ldots \to i,j\ldots)\equiv n_\psi ^{eq} n_a ^{eq}... \langle v \sigma(\psi,a\ldots \to i,j\ldots)\rangle$ notation as~\cite{Baldes:2014rda}, where $\langle v \sigma(\psi,a\ldots \to i,j\ldots)\rangle$ denotes a thermally averaged cross section, and parameterize the CP violation as 
\begin{align*}
\epsilon \equiv  \frac{\mathlarger{W (\psi,a\ldots \to i,j\ldots) - W(i,j\ldots \to \psi,a\ldots) }}{W(\psi,a\ldots \to i,j\ldots) + W(i,j\ldots \to \psi,a\ldots) }. 
\end{align*}
We define the CP symmetric reaction rate density as:
	\begin{equation}
	W_{sym} = \frac{1}{2}\Big[W(i,j\ldots \to \psi,a\ldots)+W(\psi,a\ldots \to i,j\ldots)\Big].
	\end{equation}
In general both $\epsilon_{\psi,a\ldots \to i,j\ldots}$ and $W_{sym}$ will be temperature dependent. As we will be using Maxwell-Boltzmann statistics as a (good) approximation throughout this paper, we can factor out the chemical potential from the phase space	\begin{equation}
	f_{\psi}=e^{(\mu_{\psi}-E_{\psi})/T}=e^{\mu_{\psi}/T}f_{\psi}^{eq},
	\label{chemeq}
	\end{equation}
where $f^{eq}$ refers to the equilibrium values when the chemical potential is zero, and define:
	\begin{equation}
	r_{\psi}\equiv \frac{n_{\psi}}{n_{\psi}^{eq}}=e^{\mu_{\psi}/T}.
	\end{equation}

Putting this all together, for the process $\psi + a +b + \ldots \longrightarrow i + j + \ldots$, evolving in the Friedmann-Robertson-Walker metric, the evolution of $n_{\psi}$ is given by~\cite{Kolb:1990vq}:
	\begin{equation}
  \frac{dn}{dt}+3Hn = C(\psi),
	\end{equation}
where $H$ is the Hubble expansion rate and the collision term $C(\psi)$ is:
\begin{align}
	C(\psi) 
	    =  W_{sym}\Big[r_{i}r_{j}\ldots(1-\epsilon_{\psi,a\ldots \to i,j\ldots})-r_{\psi}r_{ a}\ldots(1+\epsilon_{\psi,a\ldots \to i,j\ldots})\Big].	
	\end{align}
We compute $\langle v \sigma (\psi+a\rightarrow i + j)\rangle$ from the cross sections by using the result from \cite{Edsjo:1997bg},
\begin{equation}
\langle v \sigma (\psi+a\rightarrow i + j)\rangle=\frac{g_\psi g_a T}{8 \pi^4 n_{\psi}^{eq} n_a^{eq}}\mathlarger{\int _{(m_{\psi}+m_a)^2}^{\infty}}p_{\psi a}E_{\psi} E_a v_{\text{rel}}\sigma K_1\left(\frac{\sqrtsign s}{T}\right) ds, \label{velav}
\end{equation}
where $s$ is the square of the centre-of-mass energy, $p_{\psi a}$ is the centre-of-mass momentum of $\psi$ and $a$ and $K_i(x)$ is a modified Bessel function of the second kind of order $i$. As a note of caution, we will not use the Einstein summation convention when discussing Boltzmann equations. 
\\

The requirement that CPT and unitarity must hold imposes an important restriction on our CP violating terms. In our language the unitarity constraint is given by: \cite{Weinberg:1979bt,Dolgov:1979mz,Toussaint:1978br} 
\begin{equation}
\sum \limits_j W(i\rightarrow j)=\sum \limits_j W(j\rightarrow i)=\sum \limits_j W(i\rightarrow \overline j)=\sum \limits_j W(\overline i\rightarrow j),
\label{unitarity}
\end{equation}
where $\overline i$ is the CP transform of $i$. In general \eqref{velav} must be solved numerically, which can allow small artificial violations of unitarity to arise. To deal with these small errors, we enforce the unitarity relations \eqref{unitarity}.\\

To calculate the CP violation we use Cutkosky rules~\cite{Cutkosky:1960sp}. It should be noted that these are the cutting rules for zero temperature quantum field theory; when thermal effects dominate it is necessary to use the rules contained in~\cite{Kobes:1985kc,Bedaque:1996af}. As we will only be interested in the dominant contributions to CP violation near freeze-out, we will use the zero temperature cutting rules.\\

With these techniques, it is possible to analyse the asymmetry production.

\subsection{Interactions and CP violation}

\subsubsection{Scatterings}
We now catalogue the relevant interactions. Our CP violating scatterings are given by:
 \begin{align}
 W(Y_a L \rightarrow H_1 \phi) &\stackrel{\text{CPT}}{=} W(\phi^{*}H_1^{*} \rightarrow \overline{L} Y_a)=(1+\epsilon _{a})W_{a},\\
  W(\overline{L} Y_a \rightarrow\phi^{*}H_1^{*} ) &\stackrel{\text{CPT}}{=} W(H_1 \phi \rightarrow Y_a L)=(1-\epsilon _{a})W_{a},
 \end{align}
 where we have included an implicit sum over the three families of leptons. Because all the leptons have essentially the same mass and chemical potential, it is not really necessary to consider them as separate species except when summing the couplings for a process. These annihilations will be the main generators of the asymmetry. The relevant unitarity constraint, derived from \eqref{unitarity}, is
\begin{equation}
\epsilon_1 W_1=-\epsilon_2 W_2. 
\label{lepuni}
\end{equation}

\begin{figure}[t!]
\centering
\includegraphics[width=0.35\textwidth]{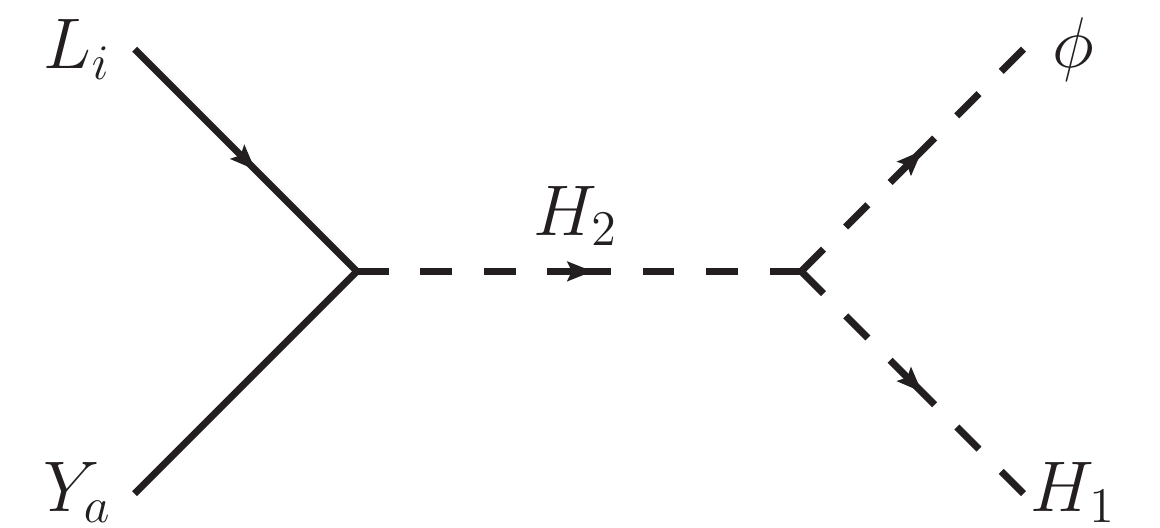} \ \ \ \ 
\includegraphics[width=0.35\textwidth]{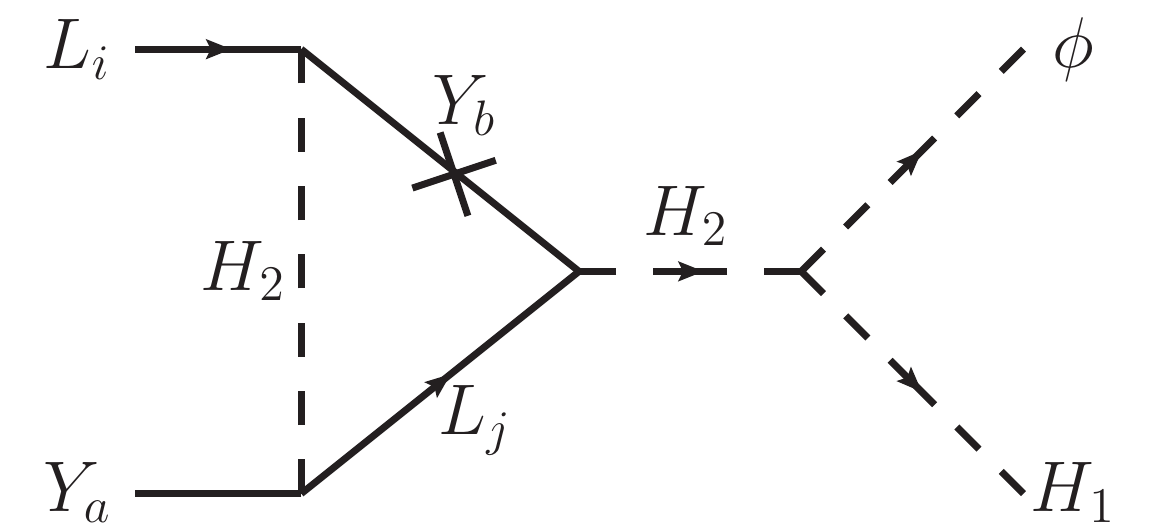}
\caption{Graphs contributing to $\epsilon_a$. The one loop graph (right) has a Majorana mass insertion.}
\label{CPmaj}
\end{figure}

 The only CP violation for the process $Y_a L \to H_1 \phi$ comes from interference between graphs in fig.~\ref{CPmaj}, involving a Majorana mass insertion. No other one loop graphs lead to a complex combination of couplings. Taking generic couplings $\lambda \sim \lambda_{ia}$ for the Lagrangian in \eqref{minlang}, the tree level cross section scales as
     \begin{equation}
     \sigma v \sim \frac{ \lambda^{2}\kappa^{2} }{ M_{H_2}^{4} }.
     \end{equation}
Denoting $\overline{\sigma}$ as the CP conjugate cross section, the CP violation at the cross section level scales as
    \begin{equation}
    (\sigma-\overline{\sigma}) v \sim \frac{ \lambda^{4}\kappa^{2} m_{Y_1}m_{Y_2}}{ M_{H_2}^{6} }.
    \end{equation}
Inserting this into \eqref{velav}, noting the momenta and energies scale as $\sqrt{s}$ and the Bessel function, which falls as $K_{1}(z) \sim \sqrt{ \frac{ \pi }{ 2z } }e^{-z}$ for $z \gtrsim 1$, imposes an effective cut-off on the integral at $\sqrt{s} \sim T$, one finds
    \begin{equation}
    W_{1} \propto \frac{ \lambda^{2}\kappa^{2} T^{6}}{ M_{H_2}^{4} }
    \end{equation}
and
    \begin{equation}
    \epsilon_{1} W_{1} \propto \frac{ \lambda^{4}\kappa^{2} m_{Y_1}m_{Y_2} T^{6}}{ M_{H_2}^{6} }.
    \end{equation}
Hence, from this dimensional analysis, it is clear there is no $T$ dependence in the CP violation, $\epsilon_{1}$ for $T\gtrsim M_{Y_2}$.\footnote{This is, of course, ignoring the implicit temperature dependence of the thermal masses, but the overall point still stands.}
We also have CP conserving scatterings, which we label
 \begin{align}
 W(Y_a H_1 \rightarrow L \phi^{*})&=T_{a},\\
 W( Y_a \phi \rightarrow H_1^{*} L_i) &=U_{a},\\
 W( Y_a L \rightarrow Y_b L) &=S_{ab},\\
 W( \overline L L \rightarrow Y_a Y_b) &=P_{ab}.
 \end{align}
While $\overline L L \rightarrow Y_a Y_b$ is not technically CP conserving, CP violation in this term only leads to a flavour asymmetry. As $Y$ is Majorana, and not ultra-relativistic, there is no method to store this asymmetry or transfer it to a non-zero $B-L$ so these terms can be safely neglected. For the t-channel graphs, no loop graph with an absorptive part is kinematically allowed and so there is no contribution to the CP violation. The expressions for the cross sections of the above processes are given in Appendix~\ref{crosssections}. \\

\subsubsection{Decays}
\begin{figure}[t!]
\centering
\includegraphics[width=0.35\textwidth]{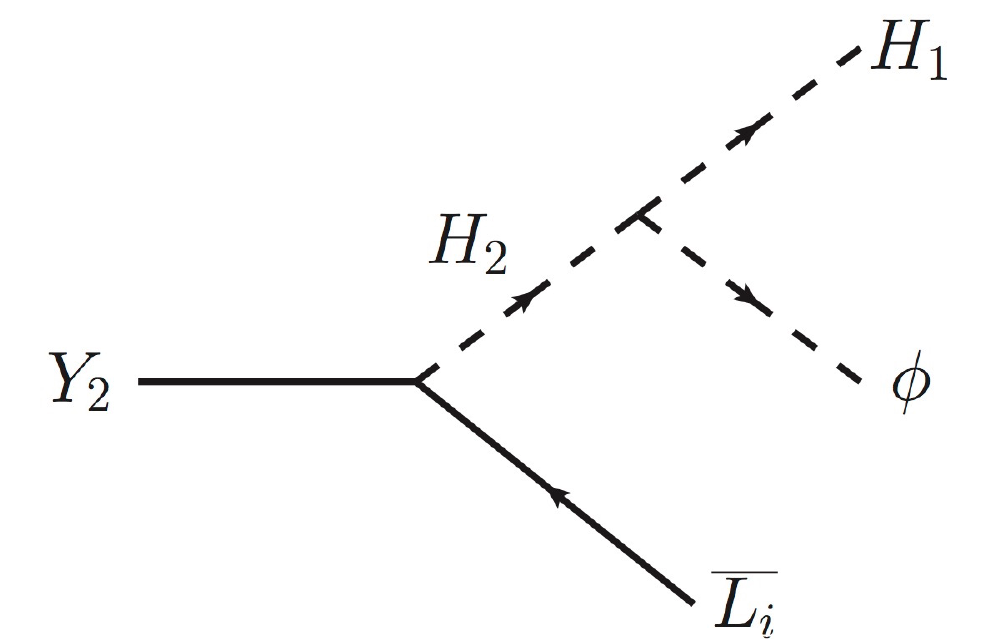} \ \ \ \ \ 
\includegraphics[width=0.4\textwidth]{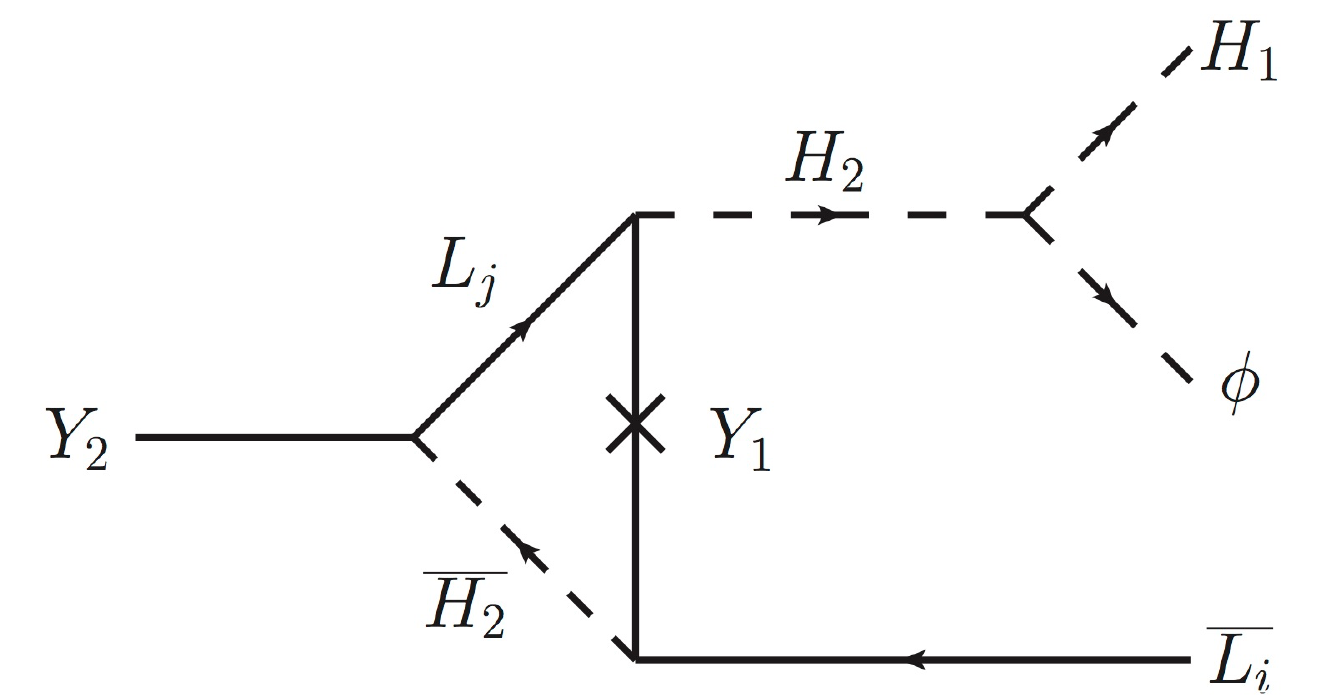}
\caption{Graphs contributing to $\epsilon_D$. The one loop graph (right) has a Majorana mass insertion.}
\label{CPmajdec}
\end{figure}

We consider the following decays of $Y_1$, $Y_2$ and $H_1$, though we are only interested in the CP violation of the $Y$ decays. We will label these decays by:
\begin{align}
\Gamma(H_1\to Y_a L \phi^{*})&=\Gamma_{H_1 \to Y_a}\label{Hdecay}\\ 
\Gamma(Y_2\to Y_1 \overline L L)&=\Gamma_{2A}\\
\Gamma(Y_2\to \overline L H_1 \phi)&=\frac{1}{2}(1+\epsilon_D)\Gamma_{2B}\\
\Gamma(Y_2\to  L H_1^{*} \phi^{*})&=\frac{1}{2}(1-\epsilon_D)\Gamma_{2B}\\
\Gamma(Y_1\to \overline L H_1 \phi)&=\Gamma(Y_1\to  L H_1^{*} \phi^{*})=\frac{1}{2}\Gamma_{1},
\end{align}
where we have parameterized the CP violation in decays by $\epsilon_D$. Note that \eqref{Hdecay} is kinematically allowed at high temperatures due to the thermal mass of $H_1$. Contributions to $\epsilon_D$ come from the graphs in fig.~\ref{CPmajdec}. We can see from this figure that $Y_1$ decays are CP conserving: CP violation would require an on shell $Y_2$ in the loop, which is kinematically forbidden. There is a further contribution to the CP violation from the $3\to3$ scattering $ \overline L H_1 \phi \to  L H_1^{*} \phi^{*}$, with the real intermediate $Y$s subtracted~\cite{Nardi:2007jp}. While the CP symmetric component of this scattering is negligible, CP violation in this process serves to cancel the CP violation in the decays at thermal equilibrium~\cite{Baldes:2014rda}. This can be seen by applying the unitarity relation \eqref{unitarity} to the state $ \overline L H_1 \phi$, to derive
\begin{equation}
\epsilon_{\overline L H_1 \phi \to  L H_1^{*} \phi^{*}}W_{\text{sym}}(\overline L H_1 \phi \to  L H_1^{*} \phi^{*})=\frac{1}{2}\epsilon_D n_{Y_2}^{eq}\Gamma_{2B},
\end{equation}
which we include in our Boltzmann equations.

\subsection{Boltzmann Equations} 
 The chemical potentials of the SM fields and $\phi$ depend only on the $B-L$ asymmetry (see Appendix~\ref{ChemicalPotentials}), so we have only three coupled differential equations to solve. These Boltzmann equations are similar to those in \cite{Baldes:2014rda}; in both cases we have two heavy Majorana particles which experience CP violation through decays and $2\to 2$ scatterings. For $Y_1$ and $Y_2$, we have:\label{boltza}
\begin{align}
\frac{dn_{Y_1}}{dt}+3H n_{Y_1}= \ &\Gamma_{1} n^{eq}_{Y_1}\Big[(r_{l} r_{\overline{H_1}} r_{\overline{\phi}}+r_{\overline {l}} r_{H_1} r_{\phi})/2 -r_{Y_1}\Big]
-\ \Gamma_{2A} n^{eq}_{Y_2}\Big[r_{l} r_{\overline{l}} r_{Y_1} -r_{Y_2}\Big] \nonumber \\
+ \ &\Gamma_{H_1 \to Y_1}  n^{eq}_{H_1}\Big[r_{H_1}+r_{\overline{H_1}}-r_{l}r_{\overline{\phi}}r_{Y_1}-r_{\overline{l}}r_{\phi}r_{Y_1}\Big]
+ \ W_1\Big[r_{H_1} r_{\phi} +r_{\overline{H_1}} r_{\overline{\phi}}-r_{Y_1}(r_{\overline{l}} + r_{l})\Big] \nonumber \\
- \ &\epsilon_1 W_1\Big[r_{H_1} r_{\phi} -r_{\overline{H_1}} r_{\overline{\phi}}+r_{Y_1}(r_{\overline{l}} - r_{l})\Big]
+ \ T_1\Big[r_{\phi} r_{\overline{l}}+ r_{\overline{\phi}} r_{l} - r_{Y_1}(r_{H_1}+r_{\overline{H_1}})\Big] \nonumber\\
+ \ &U_1\Big[r_{H_1} r_{\overline{l}}+ r_{\overline{H_1}} r_{l} - r_{Y_1}(r_{\phi}+r_{\overline{\phi}})\Big]
- \ S_{12}\Big[(r_{\overline{l}}+r_{l})(r_{Y_1}-r_{Y_2})\Big] \nonumber \\
+ \ &2P_{11}\Big[r_{\overline{l}} r_{l}-r_{Y_1}^2\Big]
+ \ P_{12}\Big[r_{\overline{l}} r_{l}-r_{Y_1}r_{Y_2}\Big].\label{B2}
\end{align}
and
\begin{align}
\frac{dn_{Y_2}}{dt}+3H n_{Y_2}= \ &\Gamma_{2B} n^{eq}_{Y_2}\Big[(r_{l} r_{\overline{H_1}} r_{\overline{\phi}}+r_{\overline l} r_{H_1} r_{\phi})/2 -r_{Y_2}\Big]
+ \ \Gamma_{2A} n^{eq}_{Y_2}\Big[r_{l} r_{\overline{l}} r_{Y_1} -r_{Y_2}\Big] \nonumber\\
+ \ &\Gamma_{H_1 \to Y_2}  n^{eq}_{H_1}\Big[r_{H_1}+r_{\overline{H_1}}-r_{l}r_{\overline{\phi}}r_{Y_2}-r_{\overline{l}}r_{\phi}r_{Y_2}\Big]
- \ \frac{1}{2}\epsilon_D \Gamma_{2B} n^{eq}_{Y_2}\Big[r_{l} r_{\overline{H_1}} r_{\overline{\phi}}-r_{\overline l} r_{H_1} r_{\phi}\Big] \nonumber \\
+ \ &W_2\Big[r_{H_1} r_{\phi} +r_{\overline{H_1}} r_{\overline{\phi}}-r_{Y_2}(r_{\overline{l}} + r_{l})\Big]
+ \ \epsilon_1 W_1\Big[r_{H_1} r_{\phi} -r_{\overline{H_1}} r_{\overline{\phi}}+r_{Y_2}(r_{\overline{l}} - r_{l})\Big] \nonumber\\
+ \ &T_2\Big[r_{\phi} r_{\overline{l}}+ r_{\overline{\phi}} r_{l} - r_{Y_2}(r_{H_1}+r_{\overline{H_1}})\Big]
+ \ U_2\Big[r_{H_1} r_{\overline{l}}+ r_{\overline{H_1}} r_{l} - r_{Y_2}(r_{\phi}+r_{\overline{\phi}})\Big] \nonumber\\
+ \ &S_{12}\Big[(r_{\overline{l}}+r_{l})(r_{Y_1}-r_{Y_2})\Big]
+ \ 2P_{22}\Big[r_{\overline{l}} r_{l}-r_{Y_2}^2\Big]
+ \ P_{12}\Big[r_{\overline{l}} r_{l}-r_{Y_1}r_{Y_2}\Big]\label{B1}
\end{align}
The Boltzmann equation for $B-L$ is:
\begin{align}
\frac{dn_{B-L}}{dt} & + 3H n_{B-L}  \nonumber  \\
& = \;  \Gamma_{1} n^{eq}_{Y_1}\Big[r_{l} r_{\overline{H_1}} r_{\overline{\phi}}-r_{\overline l} r_{H_1} r_{\phi} \Big] + \Gamma_{2B} n^{eq}_{Y_2}\Big[r_{l} r_{\overline{H_1}} r_{\overline{\phi}}-r_{\overline l} r_{H_1} r_{\phi} \Big]
 \nonumber \\
& \quad + \Gamma_{H_1 \to Y_1}  n^{eq}_{H_1}\Big[r_{H_1}-r_{\overline{H_1}}+r_{\overline{l}}r_{\phi}r_{Y_1}-r_{l}r_{\overline{\phi}}r_{Y_1}+r_{\overline{l}}r_{\phi}r_{Y_1}\Big] \nonumber \\
& \quad +  \Gamma_{H_1 \to Y_2}  n^{eq}_{H_1}\Big[r_{H_1}-r_{\overline{H_1}}+r_{\overline{l}}r_{\phi}r_{Y_2}-r_{l}r_{\overline{\phi}}r_{Y_2}+r_{\overline{l}}r_{\phi}r_{Y_2}\Big] \nonumber \\
& \quad + \epsilon_1 W_1\Big[(r_l+r_{\overline{l}})(r_{Y_2}-r_{Y_1})\Big] - \ \frac{1}{2}\epsilon_D \Gamma_{2B} n^{eq}_{Y_2}\Big[2r_{Y_2}- (r_{l} r_{\overline{H_1}} r_{\overline{\phi}}+r_{\overline l} r_{H_1} r_{\phi})\Big] \nonumber \\
& \quad + \sum\limits_{a=1,2} \Big(  W_a\Big[r_{\overline{H_1}} r_{\overline{\phi}}-r_{H_1}r_{\phi}+r_{l}r_{Y_a}-r_{\overline{l}}r_{Y_a}\Big]  + T_a\Big[r_{H_1}r_{Y_a}-r_{\overline{H_1}}r_{Y_a}+r_{\overline{\phi}}r_{\overline{l}}-r_{\phi}r_{l}\Big] \nonumber \\
 &  \quad\quad \quad \quad \quad \quad \quad \quad + U_a\Big[r_{\overline{H_1}}r_{l}-r_{H_1}r_{\overline{l}}+r_{\overline{\phi}}r_{Y_a}-r_{\phi}r_{Y_a}\Big] \Big) \nonumber \\
 \label{B3} & = \; \frac{dn_{D}}{dt}+3H n_{D}.
\end{align}
The terms with $\epsilon_1$ and $\epsilon_D$ are source terms for the asymmetry, while all other terms tend to wash out the asymmetry. We have used \eqref{lepuni} to write the Boltzmann equations solely in terms of these two CP violating terms. The decay rates which appear here are thermally averaged (see Appendix \ref{crosssections}).\\

 We can now numerically solve these to see the evolution of an asymmetry.

\subsection{Numerical Solutions}
Our numerical solutions to the Boltzmann equations are stable under changes to the precision, starting temperature (for initial temperatures $m_{H_2}\gg T\gg m_{Y_2}$) and initial conditions. We start our solutions at high temperature, where $2\to2$ scatterings can wash out any pre-existing asymmetry, and then track the evolution of $B-L$ down to the EWPT. To good approximation, at the EWPT sphalerons simply switch off, freezing the value of $B$. In general, if $Y_1$ or $Y_2$ is sufficiently long lived they may come to dominate the energy density at some time, causing the universe to become matter dominated (instead of radiation dominated as we assume). This would lead to a dilution factor due to reheating when the $Y$ decay~\cite{PhysRevD.31.681}. For the interesting regions of parameter space (those which lead to large asymmetries) this condition is not satisfied but for completeness we do include the dilution factor in our code. 
	\\
	
\begin{figure}[t!]
\centering
\includegraphics[width=0.65\textwidth]{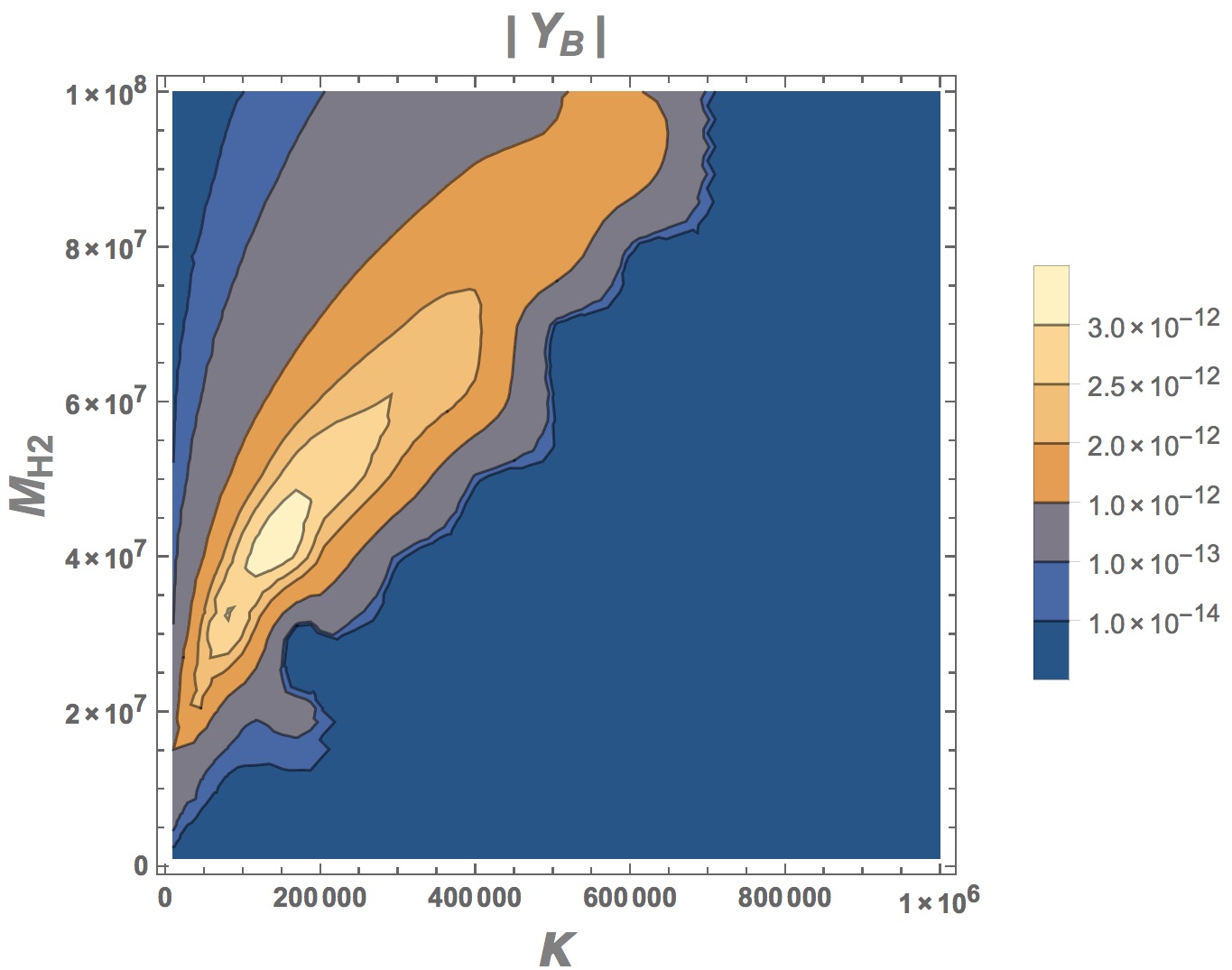}
\caption{Asymmetry formed as a function of $\kappa$ and $m_{H_2}$. There is a ridge of values where the asymmetry formed is significant, corresponding to a freeze-out temperature of order $M_{Y_2}/5$. This temperature is due to the interplay between decays and scatterings: increasing the couplings leads to $\tau_Y\lesssim t_{\text{freeze-out}}$, which forces the $Y$ to remain in equilibrium, and decreasing the couplings decreases CP violation. Maximal asymmetry corresponds to $\kappa\sim m_{Y_2}$ and $m_{H_2}\sim 10^2\times m_{Y_2}$. In this example, $m_{Y_1}=90~\text{TeV}$ and $m_{Y_2}=100~\text{TeV}$.}
\label{scan}
\end{figure}

For regimes where the two $Y$ are relatively close in mass, scatterings dominate the asymmetry, as also found in~\cite{Baldes:2014rda}. As there are two mass scales in this problem (once we choose a mass for $Y_1$ and $Y_2$), we scan over $\kappa$ and $m_{H_2}$, keeping in mind \eqref{mixing}. The asymmetry is maximised when $\kappa\sim m_{Y_1}$ and $m_{H_2}\sim 10^2\times m_{Y_1}$. Unfortunately, this model does not generate the full asymmetry required for baryogenesis; from fig.~\ref{scan} it can be seen that the maximum asymmetry is of order $Y_{B}\sim10^{-12}$.  \\

This is quite puzzling: with such similar Boltzmann equations and asymmetry production methods, how can this model fail where \cite{Baldes:2014rda} succeeded in obtaining $Y_B\sim10^{-10}$? The answer lies in the CP violation. Whereas in the neutron portal case there were graphs that gave the CP violation in scatterings the property $\epsilon(X\overline u\to d d)W(X\overline u\to d d)\propto T^2W(X\overline u\to d d)$, there is no temperature dependence in our CP violating terms. As was suggested in \cite{Baldes:2014rda}, which we will show explicitly, in the neutron portal case this temperature dependence made the scatterings relevant at a higher temperature, enhancing the asymmetry production. To obtain similar temperature dependence, we must have CP violating graphs without Majorana mass insertions. To this end we must include a second copy of the mediating scalar. In fact, this conclusion also holds for \cite{Baldes:2014rda, Baldes:2014gca}: to get the full asymmetry found in the EFTs studied it is necessary to have two heavy intermediate scalars regardless of the number of CP violating phases.

\section{Extended Neutrino Portal}
\label{leptonportalext}
\begin{figure}[t!]
\centering
\includegraphics[width=0.35\textwidth]{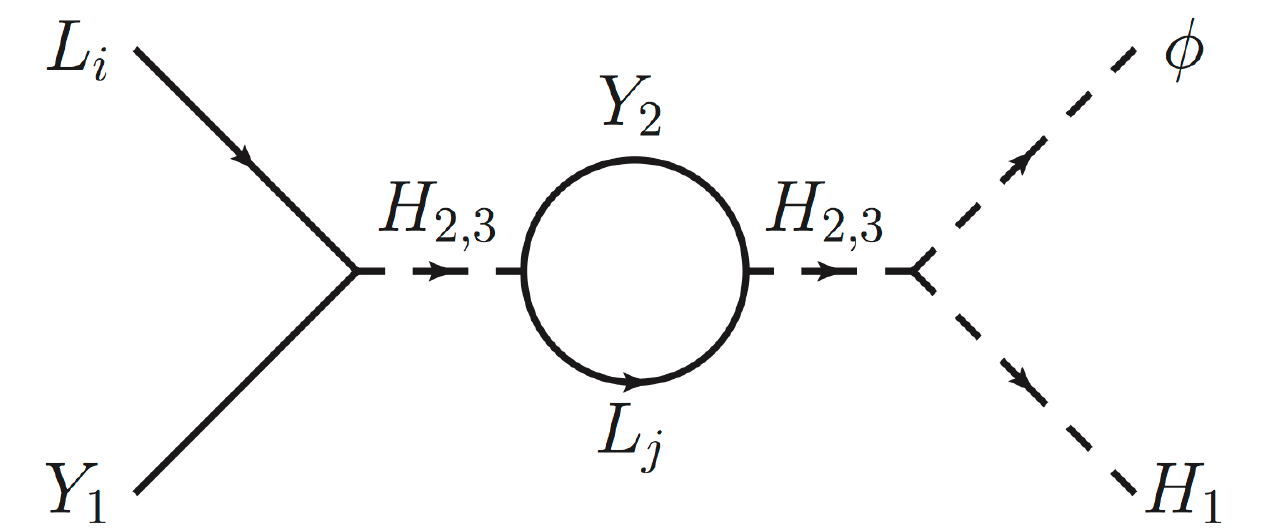} \ \ \ 
\includegraphics[width=0.35\textwidth]{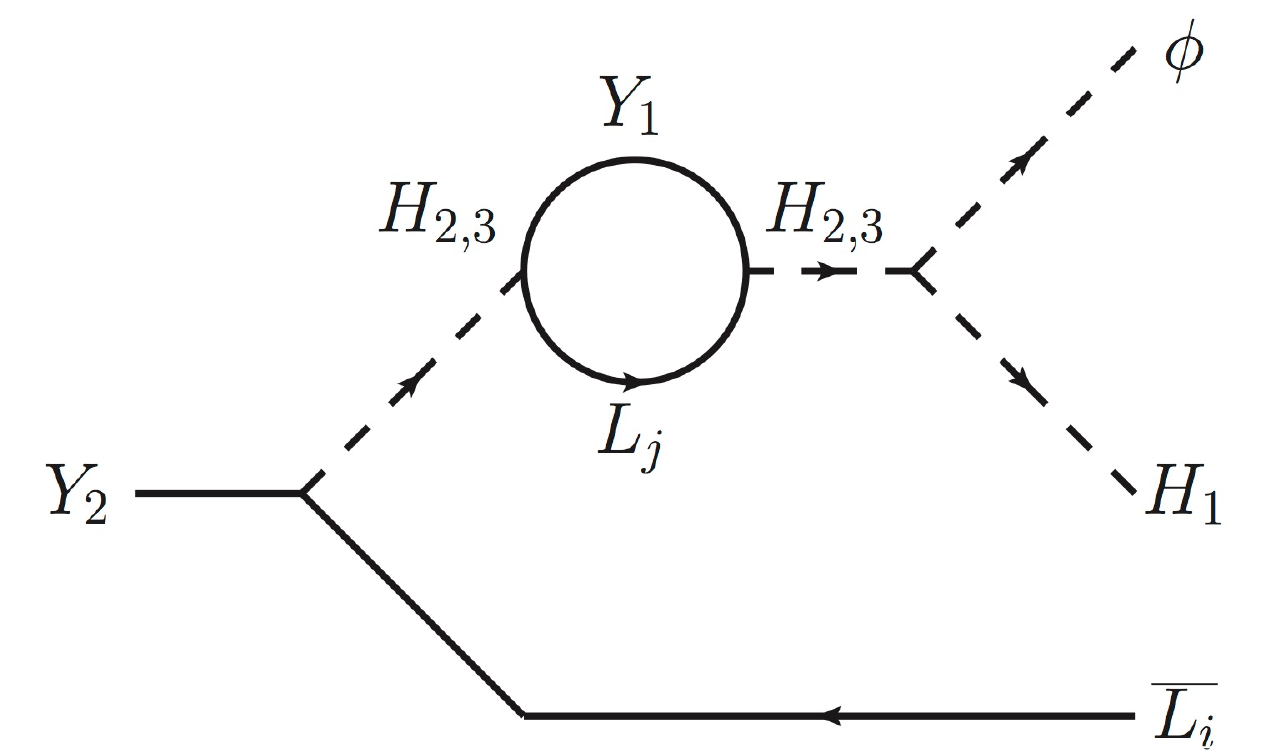}
\caption{New one loop graphs contributing to CP violation. The graph on the left (right) contributes to $\epsilon_a$ ($\epsilon_D$). Neither graph has a Majorana mass insertion.}
\label{CPext}
\end{figure}

\begin{figure}[t!]
\centering
\includegraphics[width=.6\textwidth]{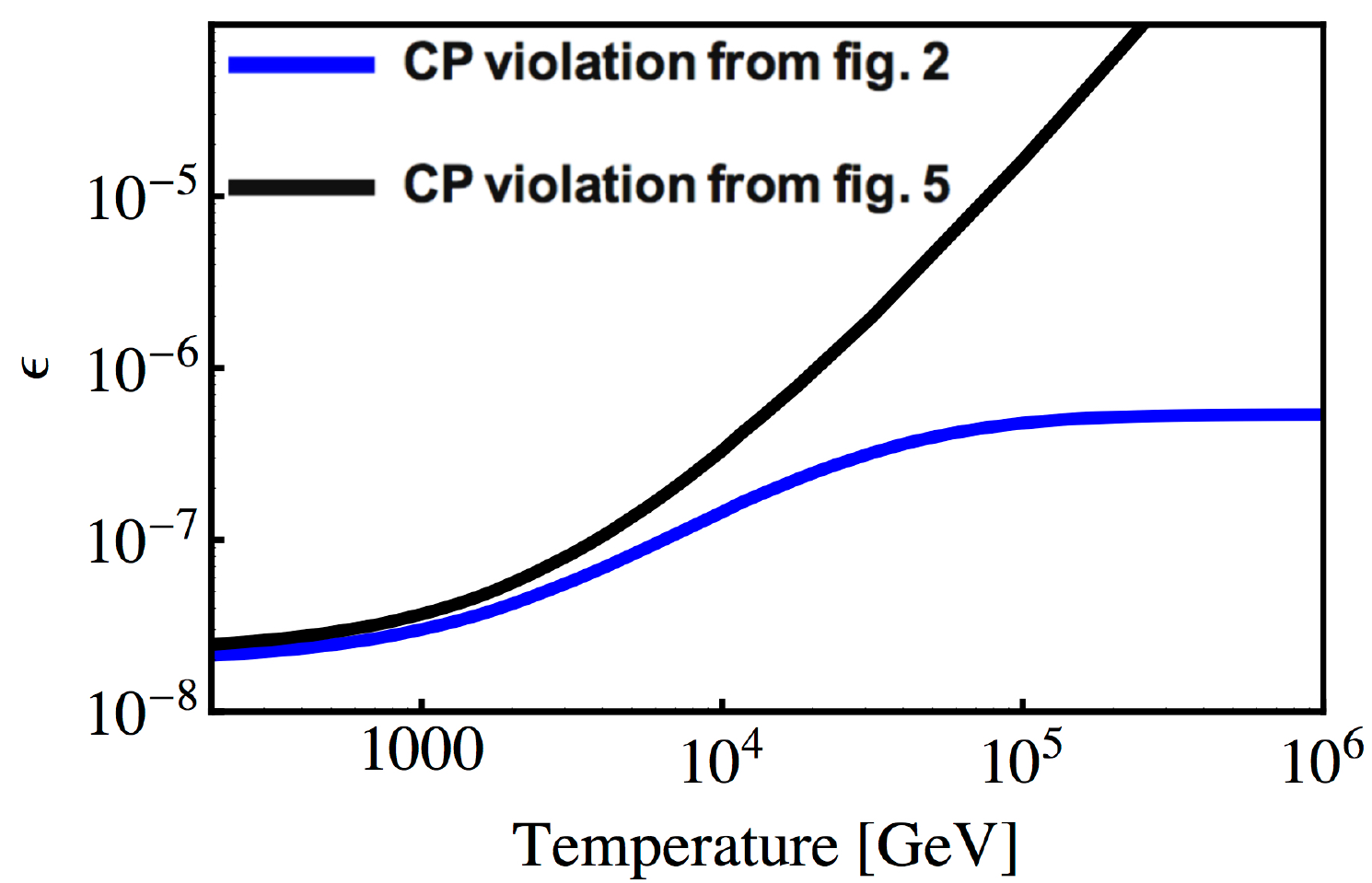} \ \ \ 

\caption{CP violation in the extended neutrino portal with $m_{Y_1}=90~\text{TeV}$ and $m_{Y_2}=100~\text{TeV}$. At high temperature the CP violation from the graphs in fig.~\ref{CPext} grows as $T^2$, and the CP violation from the graphs in fig.~\ref{CPmaj} is constant. At low temperatures both contributions become constant, and approximately equal in value. We have plotted both sources of CP violation with phases and couplings chosen to maximise their CP violation, though no parameters maximise both simultaneously. 
At $T\sim3\times 10^4~\text{GeV}$, where maximal asymmetry production occurs, the CP violation from the graphs in fig.~\ref{CPext} dominates.}
\label{CPgraph}
\end{figure}

A simple extension of the neutrino portal, which allows the generation of the observed $Y_B$, is the addition of another inert Higgs. Three-Higgs models have been explored in both an inert and general context \cite{Keus:2014jha,Ivanov:2012fp}. We will not write down the full potential, as most of the terms are irrelevant for asymmetry formation, but simply note that to satisfy electroweak precision tests it is necessary to avoid significant mixing between the two inert scalars \cite{Grimus:2007if}. The Lagrangian is now 
\begin{equation}
\Delta \mathcal L = -m_{H_p}^2|H_p|^2 - \lambda_{iap}H_p\overline L_i Y_a-\kappa_p H_1 H_p^{\dagger} \phi+ H.c,
\end{equation}
where $p=2,3$. With 14 relevant couplings there are now nine CP violating phases which, for the sake of the example, we will choose to be 
\begin{align}
\operatorname{Arg}(\lambda_{121})=&\ 0, \ \
\operatorname{Arg}(\lambda_{221})=\frac{\pi}{5}, \ \
\operatorname{Arg}(\lambda_{321})=\frac{\pi}{5}, \ \
\operatorname{Arg}(\lambda_{122})=\frac{\pi}{5}, \ \
\operatorname{Arg}(\lambda_{212})=\frac{\pi}{10} \nonumber \\
\operatorname{Arg}(\lambda_{222})=&\frac{2\pi}{5}, \ \
\operatorname{Arg}(\lambda_{312})=\frac{\pi}{10}, \ \
\operatorname{Arg}(\lambda_{322})=\frac{2\pi}{5}, \ \
\operatorname{Arg}(\kappa_2)=\frac{\pi}{3}. \label{phases}
\end{align}
We have chosen phases that avoid cancellations between the various CP violating graphs. There are now additional graphs which lead to complex couplings, involving a closed fermion loop (fig.~\ref{CPext}). As there are no Majorana mass insertions, these graphs exhibit the desired temperature dependence, $\epsilon_1W_1 \propto T^2W_1$ (see fig.~\ref{CPgraph}). This can be seen by looking at the second term in \eqref{nomajorana}, which corresponds to this new graph; relative to the first term, which is due to the original CP violation with Majorana mass insertions (fig.~\ref{CPmaj}), there are two extra factors of energy and momentum, $E$ and $p$. When velocity averaged with \eqref{velav}, at high temperatures $E\sim p\sim T$. At low temperatures the CP violation becomes constant, though due to the different combination of phases involved it is not necessarily equal to the CP violation in fig.~\ref{CPmaj}. 
There are similar contributions to the CP violation in decays, which we also include. Armed with this new CP violation, we can again solve the Boltzmann equations for the evolution of $B-L$.
\subsection{Numerical Solutions}
\begin{figure}[t!]
\centering
\includegraphics[width=.9\textwidth]{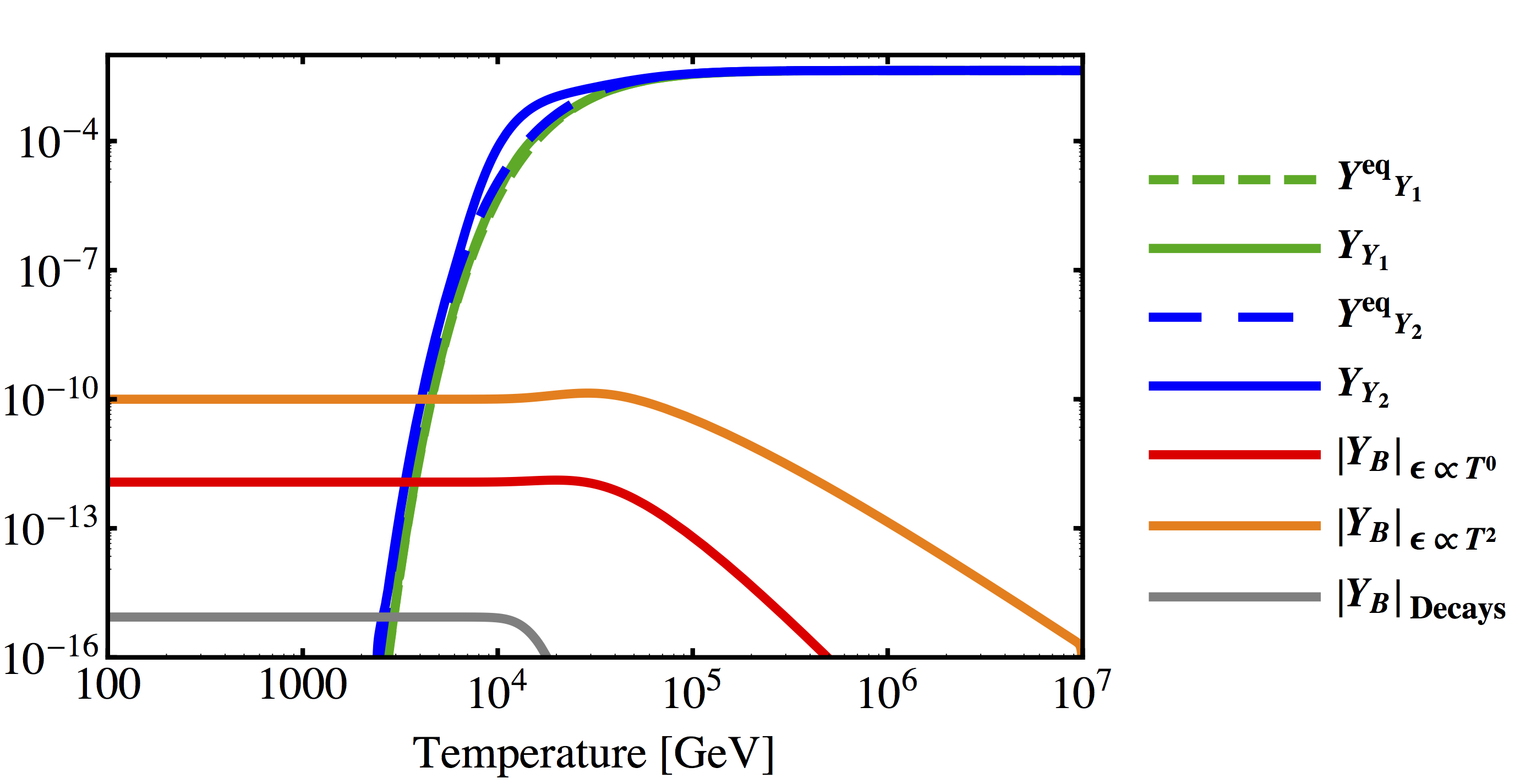}
\caption{Example solution with $m_{Y_1}=90~\text{TeV}$, $m_{Y_2}=100~\text{TeV}$. The asymmetry generated is $Y_B=1.0\times 10^{-10}$. We use $\epsilon \propto T^{0}$ and $\epsilon \propto T^2$ to refer to the CP violation in the scatterings from figures~\ref{CPmaj} and~\ref{CPext}, respectively. The asymmetry generated by the decays is negligible.}
\label{leptonportalextended}
\end{figure}

\begin{figure}[t!]
\centering
\includegraphics[width=.9\textwidth]{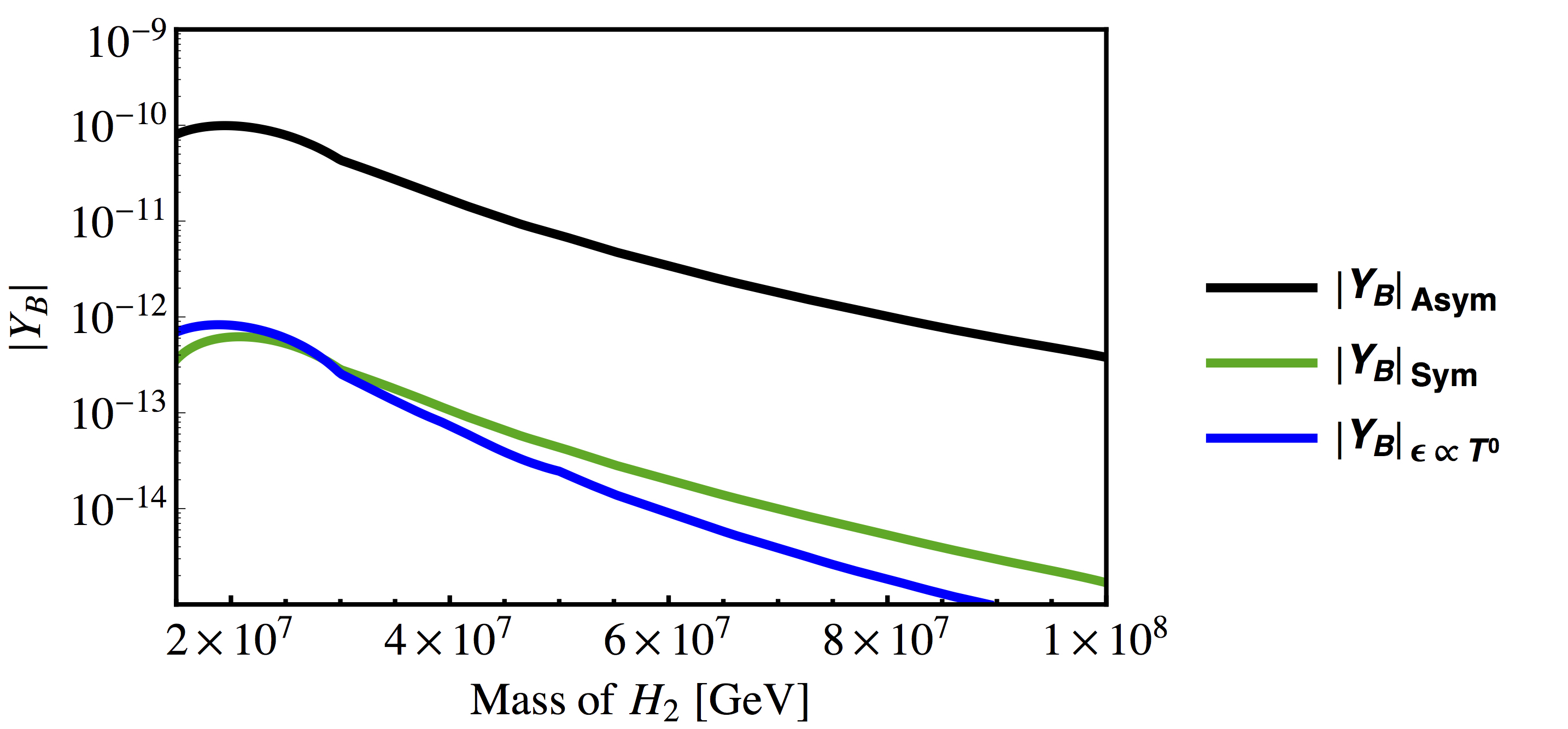}
\caption{Asymmetry formed vs $m_{H_2}$, with $m_{H_3}=2.5m_{H_2}$. For asymmetric couplings the asymmetry formed is as much as two orders of magnitude higher than when all the couplings are equal. Asymmetric couplings can avoid cancellations between the various CP violating graphs. The asymmetric couplings have been chosen for this example to maximise asymmetry production. Coupling $Y_1$ and $Y_2$ to one generation of leptons, with the couplings differing by an order of magnitude, is sufficient for a large asymmetry to form. When there is little or no discrimination between the couplings, the asymmetry formed is essentially the same as the unextended neutrino portal. If the two inert Higgs have the same masses and couplings then the temperature dependent CP violation cancels exactly.}
\label{Leptonportalasymcouple}
\end{figure}

With the temperature dependent CP violation, we see a significant increase in the asymmetry. As long as there are no significant cancellations between the various contributions to the $\epsilon_a$ the full baryon asymmetry of the universe can be generated (see fig.~\ref{leptonportalextended}). Fittingly for ADM, cancellations are minimised when the $\lambda_{iap}$ are not symmetric: it is preferable for the leptons to couple more strongly to one of the generations of the $Y$ and for not all generations of leptons to couple with the same strength. This asymmetry in the couplings does not need to be more than an order of magnitude for significant enhancement, as in fig.~\ref{Leptonportalasymcouple}. Similarly, it is preferable for $\kappa_2$ and $\kappa_3$ to be an order of magnitude different, and  $m_{H_2}$ and $m_{H_3}$ to be within an order of magnitude of each other. Comparing this to the asymmetry formed when only the mass insertion graphs are included (fig.~\ref{leptonportalextended}) we see that the asymmetry starts forming earlier, culminating in a significantly higher asymmetry. \\


Due to the similarity of the asymmetry production, Boltzmann equations and CP violation between this model and \cite{Baldes:2014rda}, it is clear that main asymmetry production in the neutron portal EFT studied in \cite{Baldes:2014rda} was also due to this temperature dependent CP violation (rather than the mass insertion diagrams which are also there in the neutron portal case). This provides compelling evidence that for scatterings to dominate over decays a heavy intermediate particle is necessary (to provide the dimensionality for temperature dependence). As many cosmological models are EFTs, this will often be satisfied. Further, multiple intermediate particles (leading to bubble graphs) seem to be a generic feature of scattering models, being necessary in not only this work but also \cite{Baldes:2014gca,Baldes:2014rda,Cui:2011ab}. \\

\section{Conclusion}
We have shown that it is possible to create a realistic UV complete model of ADM that uses scatterings to generate the asymmetry. Further, we have demonstrated explicitly the importance of temperature dependence in the CP violating terms.
\\

To do this, we introduced a new model, using an neutrino portal. By examining the $UV$ completions, we separately studied temperature independent and temperature dependent CP violation, with the latter proving to be far more significant. For future ADM model builders to see a significant asymmetry caused by $2\to2$ scatterings, quadratic temperature dependence in the CP violation will be a key feature. \\

There are many other potential operators if one is willing to allow a slightly more complicated dark sector, including variations on the WIMPy baryogenesis operators. An interesting avenue for future research would be the annihilation of the symmetric component of the $\phi$ number density; while the same scatterings cannot simultaneously create an asymmetry and annihilate the symmetric component of ADM in existing models, other minimal options can still be explored. \\
\acknowledgments
IB and AM were supported by the Commonwealth of Australia. NFB and RRV were supported
in part by the Australian Research Council.

\appendix
\section*{Appendix}
\section{Cross Sections of the Neutrino Portal}
For the sake of completeness, here we catalogue the cross sections of the case study in Sections \ref{leptonport} and \ref{leptonportalext}. $E_1$, $E_2$, $E_3$ and $E_4$ refer to the energies of the particles in the order listed. For the $2\to2$ scatterings, we will denote the initial momentum $p_i$ and the final momentum $p_f$. All cross sections are written in the centre of mass frame, except when otherwise stated.
\label{crosssections}

\subsection{Minimal Neutrino Portal}
\subsubsection{Cross sections (CP conserving component)}

We include some of the factors from \eqref{velav} to highlight the symmetry under interchange of particles.\\

\noindent\underline{$Y_aL_i\to H_1 \phi$}
\begin{equation}
p_iE_1 E_2 \sigma v=\frac{|\kappa^2 \lambda_{ia}^2|}{16\pi \sqrt s m_{H_2}^4}p_ip_f(E_1E_2+p_i^2).
\end{equation}
\underline{$L_i \phi\to Y_a H_1 $}
\begin{equation}
p_iE_1 E_2 \sigma v=\frac{|\kappa^2 \lambda_{ia}^2|}{8\pi \sqrt s m_{H_2}^4}p_ip_fE_1E_3.
\end{equation}
\underline{$L_i H_1\to Y_a \phi $}
\begin{equation}
p_iE_1 E_2 \sigma v=\frac{|\kappa^2 \lambda_{ia}^2|}{8\pi \sqrt s m_{H_2}^4}p_ip_fE_1E_3.
\end{equation}
\underline{$Y_aL_i\to Y_bL_j$}
\begin{equation}
p_iE_1 E_2 \sigma v=\frac{|\lambda_{jb}^2 \lambda_{ia}^2|}{4\pi \sqrt s m_{H_2}^4}p_ip_f\left[(E_1E_2+p_i^2)(E_3E_4+p_f^2)+E_1E_2E_3E_4+1/2p_i^2p_f^2\right].
\end{equation}
\underline{$\overline{L_i}L_j\to Y_a Y_b$}
\begin{equation}
p_iE_1 E_2 \sigma v=\frac{|\lambda_{jb}^2 \lambda_{ia}^2|+|\lambda_{ja}^2 \lambda_{ib}^2|}{4\pi \sqrt s m_{H_2}^4}p_ip_f\left[E_1E_2E_3E_4+1/2p_i^2p_f^2\right].
\end{equation}
\subsubsection{Cross sections (CP violating component)}

\noindent \underline{$Y_2L_i\to H_1 \phi$}
\begin{equation}
p_iE_1 E_2 (\sigma-\overline{\sigma}) v=-\sum\limits_j \frac{m_{Y_1}m_{Y_2}\operatorname{Im}(\lambda_{i1}\lambda_{j1}\lambda_{i2}^{*}\lambda_{j2}^{*})\kappa^2}{64\pi^2 s m_{H_2}^6}p_ip_fp_{\text{loop}}E_2E_{L_j}, \label{majcp}
\end{equation}
where $E_j$ is the energy of $L_j$ in the loop and $p_{\text{loop}}$ is the momentum of the (on shell) particles in the loop. The factor of $m_{Y_1}m_{Y_2}$ comes from projecting out the Majorana masses. CP violation in $Y_aL_i\to Y_bL_j$ is similar but only leads to flavour violation and is so neglected.\\

\subsubsection{Decays (CP conserving component)}

To calculate the three body decays, we used the usual trick of decomposing N-body phase space into a series of 2-body phase spaces. All integrals are numerically integrated. The decay rates appearing in section \ref{boltza} are in fact thermally averaged \cite{Nardi:2007jp},
\begin{equation}
\Gamma^{\text{thermal}}=\frac{K_1(m/T)}{K_2(m/T)}\Gamma.
\end{equation}
Since the relative masses of the particles changes with temperature, at different times the $Y$, $H_1$ and $\phi$ can all decay depending on the thermal masses (in particular the choice of $\lambda_4$). We will only write down the $Y$ decays as the others are similar and not particularly important. In our example solutions we chose $\lambda_4$ so that at high temperatures $H_1 \to L_i Y_a \phi$ occurs. Keeping this in mind we now catalogue the decays.\\

 \underline{$Y_a \to L_i H_1^{*} \phi^{*}$}
\begin{equation}
\Gamma=\int\limits_{(m_{\phi}+m_{H_1})^2}^{(m_{Y_a}-m_L)^2}ds \frac{4|\kappa^2 \lambda_{ia}^2|}{(2\pi)^3\sqrt{s}m_{H_2}^4m_{Y_a}^2}E_L(s)\left(\sqrt{s+p_L^2(s)}+\sqrt{m_L^2+p_L^2(s)}\right)p_L(s)p_{\phi}(s),
\end{equation}
where $E_L(s)$ is the energy of $L_i$, $p_L(s)$ [$p_{\phi}(s)$] is the momentum of $L_i$ [$\phi$] in the centre-of-momentum frame [rest frame of the mediating particle $H_2$] and $s$ is $p_{\phi}^{\mu}p_{H_1 \mu}$. When we say the rest frame of $H_2$, we mean a fictitious on-shell particle with mass $\sqrt s$ in place of $H_2$. \\

\noindent \underline{$Y_2 \to \overline{L_i} Y_1  L_j$}
\begin{align}
\Gamma=&\int\limits_{(m_{L_j}+m_{Y_1})^2}^{(m_{Y_2}-m_{L_i})^2}ds \frac{4(|\lambda_{\lambda_{i1}^2 j2}^2|+|\lambda_{j1}^2 \lambda_{i2}^2|)}{(2\pi)^3\sqrt{s}m_{H_2}^4m_{Y_2}^2}E_{L_i}(s)\left(\sqrt{s+p_{L_i}^2(s)}+\sqrt{m_{L_i}^2+p_{L_i}^2(s)}\right)\\
\times& p_{L_i}(s)p_{L_j}(s)\left(E_{L_j}(s)E_{Y_1}(s)+p_{L_j}^2(s)\right),\nonumber
\end{align}
where $E_{L_{i(j) }}(s)$ and $p_{L_{i(j) }}(s)$ are the energy and momentum of $L_{i~(j) }$, respectively, and $s$ is $p_{Y_2}^{\mu}p_{L_j \mu}$. Note that $p_{L_{j}}(s)$ is in the rest frame of $H_2$ and all others are in the rest frame of $H_2$.\\

\subsubsection{Decays (CP violating component)}

 \underline{$Y_2 \to L_i H_1^{*} \phi^{*}$}
\begin{equation}
\Gamma-\overline \Gamma= \int\limits_{Low}^{(m_{Y_2}-m_L)^2}ds\sum \limits_j \frac{m_{Y_1}\operatorname{Im}(\lambda_{i1}\lambda_{j1}\lambda_{i2}^{*}\lambda_{j2}^{*})\kappa^2}{(2\pi)^4s^{3/2}m_{H_2}^6m_{Y_2}}E_{L_J}(s)\left(m_{Y_2}^2-s-m_{L_i}^2\right)p_{L_i}(s)p_{L_j}(s)p_{\phi}(s),
\end{equation}
where $s$ is $p_{L_i}^{\mu}p_{H_1 \mu}$ and $\text{Low}\equiv\operatorname{Max}[(m_{\phi}+m_{H_1})^2,(m_{Y_1}+m_{L_j})^2]$. The kinematics of $L_i$ were written in the centre-of-momentum frame, and the other particles kinematics are in the rest frame of $H_2$. 

\subsection{Extended Neutrino Portal}
When a second inert Higgs doublet is added, $H_3$, we get many more diagrams contributing to the various processes. For most processes, these diagrams are simply allowing the intermediate particle to be $H_2$ or $H_3$, with some interference terms. Due to the triviality of the extension, we do not write these down. But there are important new contributions to the CP violation in both the decays and scatterings. The new cross section is:\\

\noindent \underline{$Y_2L_i\to H_1 \phi$}
\begin{align}
p_iE_1 E_2 (\sigma-\overline{\sigma}) v=&-\sum\limits_{jprs} \frac{m_{Y_1}m_{Y_2}\operatorname{Im}(\lambda_{i1r}\lambda_{j1s}\lambda_{i2p}^{*}\lambda_{j2r}^{*}\kappa_p^{*}\kappa_s)}{64\pi^2 s m_{H_p}^2m_{H_r}^2m_{H_s}^2}p_ip_fp_{\text{loop}} E_2E_{L_j} \\ &+\sum\limits_{jprs}\frac{\operatorname{Im}(\lambda_{j1r}^{*}\lambda_{j1s}\lambda_{i2p}^{*}\lambda_{i2r}\kappa_p^{*}\kappa_s)}{32\pi^2 s m_{H_p}^2m_{H_r}^2m_{H_s}^2}p_ip_fp_{\text{loop}}\left(E_1E_2+p_i^2\right)\left(E_{L_j}E_{Y_1}+p_{loop}^2\right),\nonumber 
\label{nomajorana}
\end{align}
where $p,~r,~s\in 2,~3$ and $E_{Y_1}$ is the energy of $Y_1$ in the loop. The first term is essentially the same as \eqref{majcp}, and is due to an extended analogue of fig. \ref{CPmaj}. The second term is due to fig. \ref{CPext}: as there are no Majorana mass insertions we have a more complicated function of energy and momenta in place of the Majorana masses. We note that if all the masses and couplings are the same for $H_2$ and $H_3$ the second term cancels, and CP violation is reduced to the minimal neutrino portal. The contributions to the decays are given by: \\

\noindent \underline{$Y_2 \to L_i H_1^{*} \phi^{*}$}
\begin{align}
\Gamma-\overline \Gamma=&- \int\limits_{Low}^{(m_{Y_2}-m_L)^2}ds \sum \limits_{jprs}\frac{m_{Y_1}\operatorname{Im}(\lambda_{i1r}\lambda_{j1s}\lambda_{i2p}^{*}\lambda_{j2r}^{*}\kappa_p^{*}\kappa_s)}{(2\pi)^4s^{3/2}m_{H_p}^2m_{H_r}^2m_{H_s}^2m_{Y_2}}E_{L_J}(s)\left(m_{Y_2}^2-s-m_{L_i}^2\right)\nonumber \\
&\times p_{L_i}(s)p_{L_j}(s)p_{\phi}(s) \nonumber\\
&+\int\limits_{Low}^{(m_{Y_2}-m_L)^2}ds \sum \limits_{jprs}\frac{2\operatorname{Im}(\lambda_{j1r}^{*}\lambda_{j1s}\lambda_{i2p}^{*}\lambda_{i2r}\kappa_p^{*}\kappa_s)}{(2\pi)^4s^{3/2}m_{H_p}^2m_{H_r}^2m_{H_s}^2m_{Y_2}^2}E_{L_J}(s)\left(\sqrt{s+p_{L_i}^2(s)}+\sqrt{m_{L_i}^2+p_{L_i}^2(s)}\right)\nonumber \\
&\times\left(E_{L_j}E_{Y_1}+p_{loop}^2\right)p_{L_i}(s)p_{L_j}(s)p_{\phi}(s).
\end{align}
Similar to the CP violation in the scatterings, in the limit of degenerate couplings and masses the second term disappears. This completes the cataloguing of the cross sections used to solve the Boltzmann equations in section \ref{boltza}.

\section{Thermal Masses}
\label{ThermalMasses}
While throughout this paper most thermal field theoretic effects are neglected, there is one that must be included. At high temperatures particles can gain an effective mass through interactions with the plasma. For $H_1$, $L$ and $\phi$, these thermal masses are kinematically significant. $H_2$ is too massive to appear in the plasma at the temperatures we consider and so does not acquire a thermal mass, nor does it contribute to thermal masses. Similarly, as the $Y$ only interact via $H_2$ they also do not acquire a thermal mass. The main effect of thermal masses is to change the kinematics; it has been shown that effects such as the apparent breaking of chiral symmetry can be neglected to good enough approximation for our purposes \cite{Kiessig:2009cm,Giudice:2003jh,Kiessig:2011ga}. The thermal masses are given by: \cite{Kiessig:2011ga,Katz:2014bha}
\begin{align}
m_{H_1}^2=&\left(\frac{3}{16}g_2^2+\frac{1}{16}g_Y^2+\frac{1}{4}y_t^2+\frac{1}{2}\lambda_{H_1}\right)T^2,\\
m_{L}^2=&\left(\frac{3}{32}g_2^2+\frac{1}{32}g_Y^2\right)T^2,\\
m_{\phi}^2=&\frac{\lambda_4 T^2}{2},
\end{align}
where $g_2$ is the coupling constant of $SU(2)_L$ and $g_Y$ is the coupling constant of $U(1)_Y$  in the SM. We have neglected all SM Yukawa couplings except to the top quark, as well as contributions from $\lambda_3$. In our solutions we use $m_{H_1}=0.71T$, $m_{\phi}=0.59T$ and $m_L=0.19T$. In general, the asymmetry formed increases for large thermal masses. \\

Interestingly, as the $Y$ do not acquire a thermal mass, at temperatures above $1.4m_{Y_a}$ the SM Higgs can decay into $Y$ rather than the other way round (depending on the thermal mass of $\phi$). While there could be concerns about the CP violation of these decays, these are ineffective as $H_1$ is in equilibrium. This is in agreement both with ansatz calculations of the CP violation and the results from \cite{Giudice:2003jh}.  

\section{Chemical Potentials}
\label{ChemicalPotentials}
Since $\phi$, as well as the leptons, will be kept close to equilibrium above the EWPT it is unnecessary to have a Boltzmann equation for each particle species. Rather, we can solve our Boltzmann equations with the chemical potentials of these species, using \eqref{chemeq} to obtain their number densities. This is where the advantage of using Maxwell-Boltzmann statistics lies: we can separate out the chemical potentials. Here we are not considering processes (or the relevant chemical potentials) below the EWPT. Our task is made simpler by the fact that all our chemical potentials can be written in terms of $\mu_{\phi}$. As our model is similar to standard treatments of $B$ and $L$ violation, such as leptogenesis, we can borrow the chemical potentials from \cite{Buchmuller:2000as}, and just note that the segregation of $B-L$ into $\phi$ is the only source of $B-L$ violation. At the temperatures we will be considering, the population of $H_2$ is negligible and so does not affect the chemical potentials. By making the replacement
\begin{equation}
B-L=\mu_{\phi},
\end{equation}
we obtain the chemical potentials:
\begin{align}
\mu_L&=\frac{-7}{79}\mu_{\phi}\\
\mu_{H_1}&=\frac{4}{79}\mu_{\phi}.
\end{align}
These can be related to the asymmetry generated in $\phi$ by using 
\begin{equation}
n_{\phi}-n_{\overline{\phi}}=\frac{T^2}{3}\mu_{\phi}.
\end{equation}
To write this all in terms of the baryon asymmetry, we use 
\begin{equation}
 B=\frac{28}{79}\mu_{\phi}.
\end{equation}

\section{Extensions of the Neutron Portal}
\label{neutronportal}
\subsection{$2\to3$ scatterings}
We have considered the simplest gauge invariant combination of $B-L$ carrying SM particles, $\overline L H$, but what about the second simplest? The next lowest dimensional gauge singlet $B-L$ combination of SM particles is $udd$, the neutron portal. The neutron portal's main issue is stability - the dark matter tends to decay. There are two ways to stabilise the neutron portal: temperature dependent couplings, as suggested in~\cite{Farrar:2005zd}, or a symmetry. Both of these approaches require additional particle degrees of freedom. If we wish to impose a new symmetry, we must move beyond $2\to 2$ scatterings. The essential neutron portal operator is 
\begin{equation}
\overline X u_R \overline{d_R^c} d_R, \label{neut}
\end{equation}
where $X$ is Dirac. It also possible to use left handed quarks, or for $X$ to be Lorentz contracted with down type quarks, but there are no differences relevant to this discussion. \\

Following the method of Section~\ref{alternatives}, the simplest way to make a stable version of the neutron portal is to introduce the effective operator
\begin{equation}
g\overline X u_R \overline{d_R^c} d_R \sigma, \label{stable}
\end{equation}
where $X$ is a Majorana fermion and $\sigma$ is a complex SM singlet scalar.\footnote{An alternative ADM scenario, using the neutron portal and CP violating decays, can be found in \cite{Davoudiasl:2010am}.} As before, to satisfy \eqref{unitarity} we require two generations of $X$. We will only consider one generation of quarks: as they remain in equilibrium any CP violation from having multiple generations will not change our discussion below.
\\

\begin{figure}[t!]
\centering
 \includegraphics[width=350pt]{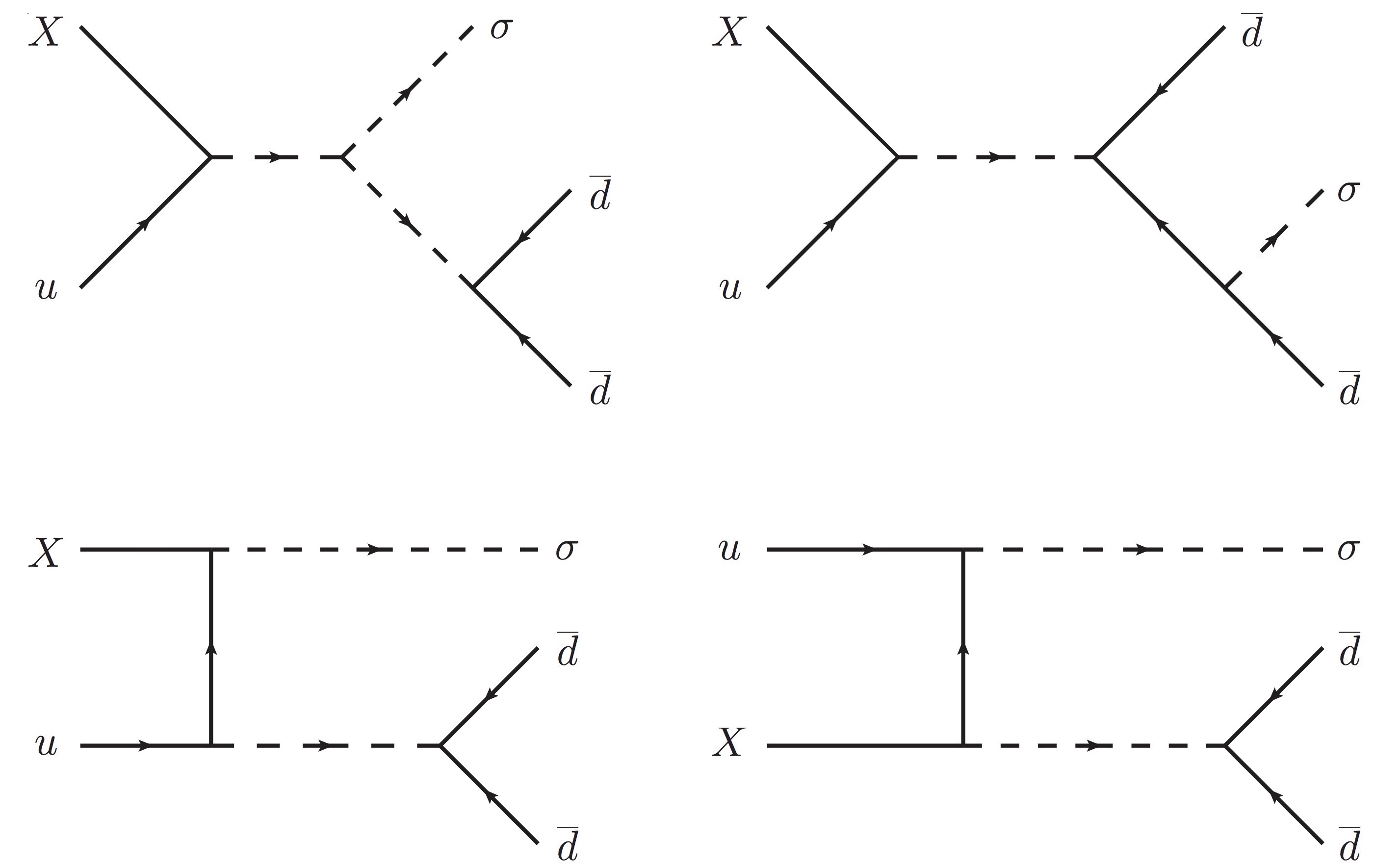} \ \ \ \
\caption{Tree level UV completions of the stable neutron portal operator \eqref{stable}.}
\label{fig:3}
\end{figure}

\begin{figure}[t!]
\centering
\includegraphics[width=0.35\textwidth]{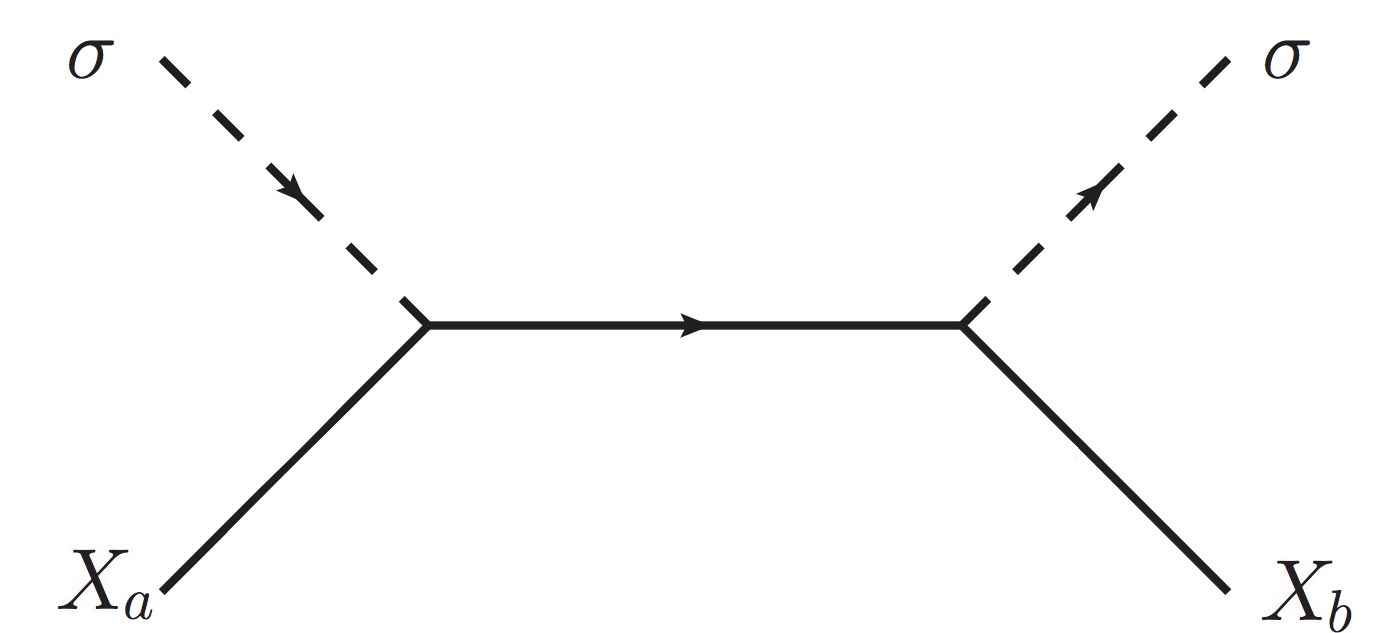} \ \ 
\includegraphics[width=0.35\textwidth]{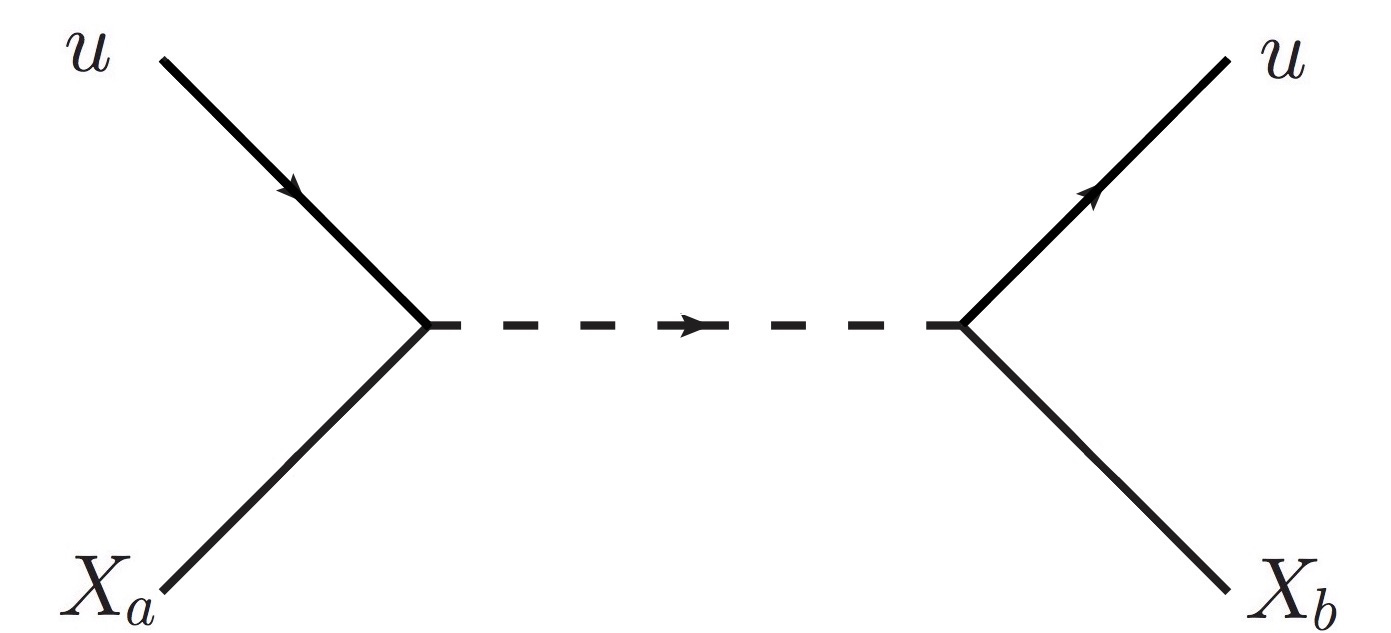}

\caption{Flavour changing scatterings.}
\label{fig:4}
\end{figure}

There are multiple ways to open up this effective operator, shown in fig.~\ref{fig:3}, but the only simple UV completion capable of generating an asymmetry is the last. In all others, there are rapid $2 \rightarrow 2$ flavour changing scatterings (fig.~\ref{fig:4}), which are problematic. As the $2\to2$ scatterings are mediated by only one heavy intermediate scalar, they dominate over the $2\rightarrow3$ scatterings. As CP violating effects involving the $2 \rightarrow 3$ scatterings must be as large as possible during freeze-out the delay of thermal non-equilibrium due to $2\to2$ scatterings results in a serious suppression of the final asymmetry. \\

If the collision terms for $u$ and the $X$ are given by
\begin{align}
W(u +  X_1 \rightarrow \overline d +\overline d +\overline {\sigma})&=(1+\epsilon_1)W_1,\nonumber \\
W(u + X_2 \rightarrow \overline d +\overline d +\overline {\sigma})&=(1+\epsilon_2)W_2,\\
W(u +  X_1 \rightarrow u +  X_2)&=(1+\epsilon_3)W_3,\nonumber
\end{align}
then the relevant unitarity constraints are given by
\begin{align}
\epsilon_1W_1+\epsilon_2W_2&=0,\nonumber \\
\epsilon_1W_1+\epsilon_3W_3&=0, \\ \nonumber
\epsilon_2W_2-\epsilon_3W_3&=0.
\end{align}
From this, it is clear that for any of these process to violate CP, flavour changing scatterings must exist. Further, to get significant CP violation without delaying the departure from thermal equilibrium these scatterings must be comparable in size to the $2 \to 3$ scatterings. This argument holds for any of the particles $X$ could scatter with; at least one flavour changing process must exist. As in this last UV completion there are no tree level scatterings to accomplish this, another process must be added.\footnote{Mediators for this process are contained in the other UV completions.}\\

It seems that the only way to obtain a working ADM neutron portal using $2\to3$ scatterings is to have correlated couplings, which is not obviously an improvement over temperature dependent couplings. Any full theory of $2\to 3$ scatterings must explain why two seemingly unrelated processes conspire to allow an asymmetry to be formed. While the neutron portal is the second simplest SM gauge singlet operator involving baryon or lepton number, it actually requires a complex description. We must look elsewhere for a generic method for ADM via scatterings.

\subsection{What about the symmetric component?}
\label{whataboutsym}
As discussed in Section \ref{leptonport}, no existing models can use the same process to annihilate the symmetric component and generate an asymmetry but is it possible to design one that does? Any model satisfying this condition must have four properties:
\begin{enumerate}
\item{The particles going out-of-equilibrium must be stable.}
\item{The process must freeze-out late enough for the symmetric component to be annihilated, $T_{\text{freeze-out}}\lesssim m_{DM}/25$.\label{freeze}}
\item{Dark matter must couple to quarks, as a lepton asymmetry cannot be reprocessed below $T_{EWPT}$ and ADM requires that $m_{DM}$ is below the EW scale.}
\item{CP violation must be significant at freeze-out, at least $10^{-6}$ (a conservative lower bound by analogy with \cite{Bernal:2012gv}, which used kinematic suppression of the washout to enhance the asymmetry formed).\label{cpviol}}
\end{enumerate}
As there are strict bounds on light non-SM coloured particles \cite{Aad:2014wea}, DM must couple to a gauge singlet combination of quarks carrying baryon number, so some extension of the neutron portal is necessary. At least one of the mediators for any interaction involving the neutron portal must be coloured, making it potentially detectable at the LHC. We can derive an upper bound on $m_{\text{mediators}}$ from properties \ref{freeze} and \ref{cpviol}. As dark matter is very light (order GeV), this is in tension with collider searches, which require that new coloured particles are heavier than about a TeV (for gluino-like particles) and 440 GeV (for single light flavoured squark-like particles)~\cite{Aad:2014wea}. Models with small mass splittings can weaken these bounds, however as we are considering ADM the dark matter will always be significantly lighter than the coloured mediators. \\

 The simplest model which satisfies these conditions is \eqref{stable},\footnote{If we make $X$ Dirac, rather than Majorana, and kinematically disallow decays dark matter will be an admixture of $X$ and $\sigma$, similar to \cite{Davoudiasl:2010am}.} so we can get the most forgiving bounds from it. More complicated extensions of the neutron portal will in general have tighter bounds, as additional mediators suppress the interaction. To simplify matters we will consider all mediators to have the same mass; light SM gauge singlet mediators tend to lead to rapid $D$ conserving scatterings amongst the dark sector particles.  We estimate the freeze-out temperature by comparing the interaction rate $\Gamma$ to the Hubble time: freeze-out occurs when $\Gamma\sim H$. These results agree with the numerical calculations of the similar neutrino portal model. Requiring that freeze-out occurs late enough (point \ref{freeze} above) gives
\begin{equation}
 m_{\text{mediator}}\lesssim150\times  \lambda~\text{GeV},
\end{equation}
where $\lambda$ is the couplings strength, assuming all couplings are the same. The CP violation at low temperatures is bounded by a loop factor $\frac {m_{\text{DM}}^3}{\pi m_{\text{mediator}}^3}$. From \ref{cpviol} we get \begin{equation}
 m_{\text{mediator}}\lesssim200\times  \lambda~\text{GeV}.
 \end{equation}
 While limits on coloured particles are model dependent, coloured particles of order $150$ GeV would almost certainly have been seen at the LHC. The squark limits can be avoided for $\lambda \gtrsim 3$, but this raises the issue of perturbativity and low energy Landau poles. So strong a Yukawa interaction between quarks and the mediating particles means that there would be a very strong detection possibility via, for example, monojet searches. This is an interesting avenue for future work.\\

Unfortunately it seems that no operator allows one to use a single scattering process to annihilate the symmetric component and generate an asymmetry through asymmetric freeze-out without using couplings greater than one. Further, this requires at least $2\to3$ scatterings, so for the same process to annihilate the symmetric component and generate an asymmetry simultaneously, it must rely on a conspiracy with flavour violating scatterings.

\section{Other $2\to2$ ADM Scattering Operators}
\label{otherops}
As was shown in Appendix \ref{neutronportal}, $2\to2$ scatterings have significant advantages over $2\to3$ scatterings. Other than the neutrino portal, what other $2\to2$ scatterings are possible? If the SM particles involved are not required to form a gauge singlet then there are dozens of operators that can be used. Due to the sheer number of operators (many being only trivially different) an exhaustive list would be exhausting both to the reader and the authors. Instead, we will discuss the possibilities using an example.\\

For the example, consider WIMPy baryogenesis~\cite{Cui:2011ab}. WIMPy baryogenesis also uses CP violating scatterings to generate a baryon asymmetry, albeit using the "WIMP miracle" instead of an asymmetry to set the relic abundance of dark matter. Despite this, it is possible to tweak WIMPy baryogenesis so that it becomes a theory of ADM. Fortuitously, the different ways of modifying WIMPy baryogenesis reveal all the salient model building concerns of $2\to2$ ADM scattering operators. For stock WIMPy baryogenesis, the relevant additions to the Lagrangian are~\cite{Cui:2011ab} 
 \begin{equation}
 \Delta \mathcal L= g\overline L \psi\overline{X}^cX + \lambda H\overline\psi f+ H.c,
 \end{equation}
 where $X$ and $f$ are singlet Dirac fermions, the dark matter, and $\psi$ carries the same quantum numbers as $L$. In the original model, the $X$ freeze-out, not carrying any baryon number, behave as WIMPs. CP violation in these scatterings stores lepton number in the $\psi$. The $\psi$s (which are heavier than $X$) decay via $\psi \rightarrow H^{*} f$. Dangerous decays giving lepton number back to the visible sector are forbidden by a $\mathds{Z}_4$ symmetry. Of course, there are other WIMPy baryogenesis models with, for example, $X$ coupling to quarks but these give a similar story. There are two different approaches to turning the WIMPy baryogenesis operator into a model of ADM, corresponding to the two broad categories of $2\to2$ ADM scattering operators.\\
 
The first approach corresponds to a class of theories similar to the toy model of~\cite{Baldes:2014gca}.\footnote{In a sense, the toy model itself can be considered a highly non-minimal realistic model, using no SM particles. It can still be made realistic by making, for example, the particles that go out-of-equilibrium decay into SM particles.} In this first class of theories, particles that store an asymmetry go out-of-equilibrium and then decay into lighter particles, transferring the asymmetry. This decay is required to allow the annihilation of the symmetric component. 
To accomplish this, and to get a tighter relationship between dark and visible matter, we can do away with $X$. There is already a conserved U(1) symmetry between $L$ and $\psi$, so the process \begin{equation} 
LL\to \psi \psi 
 \end{equation}
could create an asymmetry through asymmetric freeze-out, with $\psi$ going out of equilibrium. The asymmetry stored in $\psi$ is transferred to $f$ as in WIMPy baryogenesis, but now dark matter consists solely of $f$. In place of $X$, a mechanism to annihilate the symmetric component of $f$ is necessary but for essentially the same degrees of freedom we now have $\Omega_{DM} \simeq5\Omega_{VM}$ naturally. \\

In the second approach, which is similar to the neutrino portal model and \cite{Baldes:2014rda}, particles that do not carry $B-L-D$ depart from equilibrium. In the WIMPy baryogenesis scenario, we can replace one of the $X$s in this operator with a Majorana fermion Y, giving
\begin{equation}
  \Delta \mathcal L=  g\overline L \psi\overline{X}^cY + \lambda H\overline\psi f+ H.c.
\end{equation}
Now $X$ is free to be light, and to remain in thermal equilibrium. As in the neutrino portal model, the Majorana fermion $Y$ departs from equilibrium, and subsequently decays. Unlike WIMPy baryogenesis, we insist that $X$ carry a baryon number equivalent. This leads to two component dark matter, consisting of $X$ and $f$. In order to ensure the stability of $f$ we introduce a $\mathds{Z}_2$ symmetry, with $Y$, $\psi$ and $f$ being the negative parity states. If $X$ is lighter then $f$, $N_X-(N_f+N_{\psi})$ conservation ensures the stability of $X$. To annihilate the symmetric component, it is possible to gauge $\text{U}(1)_{N_X-(N_f+N_{\psi})}$. This should produce a similar result to the neutrino portal as the Boltzmann equations and types of CP violation are quite similar, though it is non-minimal as it only involves one SM particle.\\

All other operators using this mechanism of asymmetric freeze-out should fall broadly into one of these two categories. While the neutrino portal is the minimal case, adding complexity to the dark sector can lead to some potentially interesting phenomenology, such as dark forces.

\label{Bibliography}

\footnotesize

\renewcommand\bibsection{\section*{Bibliography}}
\bibliographystyle{utphys} 
\providecommand{\href}[2]{#2}\begingroup\raggedright\endgroup

\end{document}